\documentclass[11pt,a4paper]{article}
\usepackage[utf8]{inputenc}
\usepackage[T1]{fontenc}
\usepackage{lmodern}
\usepackage{microtype}
\usepackage[margin=2.4cm]{geometry}
\usepackage{graphicx}
\usepackage{pdflscape}
\usepackage{amsmath,amssymb}
\usepackage{array,booktabs,longtable}
\usepackage[font=small,labelformat=empty]{caption}
\usepackage[shortlabels]{enumitem}
\usepackage[round]{natbib}
\usepackage{xurl}
\usepackage[hidelinks]{hyperref}
\newcolumntype{L}[1]{>{\raggedright\arraybackslash}p{\dimexpr #1\linewidth-2\tabcolsep\relax}}
\newcommand{\tsrc}{{\footnotesize\noindent Source: Authors' own work\par}\medskip}
\hypersetup{pdftitle={Technology-Enabled Fuel Supply Chain Resilience: An Integrated Information System for Real-Time Monitoring and Decision Support}, pdfauthor={Mohamed Loutis; Esma A\"{\i}meur}}
\title{Technology-Enabled Fuel Supply Chain Resilience: An Integrated Information System for Real-Time Monitoring and Decision Support}
\author{Mohamed Loutis\thanks{Corresponding author.}\\{\small Sorbonne Université, Paris, Île-de-France, France}\\{\small mohamed.loutis@etu.sorbonne-universite.fr; ORCID 0009-0001-6491-9193} \and Esma Aïmeur\\{\small Université de Montréal, Montreal, Quebec, Canada}\\{\small esma.aimeur@umontreal.ca; ORCID 0000-0001-7414-5454}}
\date{}
\begin{document}
\maketitle

\begin{abstract}
\noindent\textbf{Purpose} -- To evaluate and extend the transactional information system described in an earlier fuel logistics study, focusing on data reliability, monitoring and decision support.\par\smallskip
\noindent\textbf{Design/methodology/approach} -- An author-led re-evaluation examined available Delphi/SQL code and tested a reconstructed source-derived baseline (R0) against a revised SQL Server artefact (R1) in twelve scenarios. Synthetic transactions and public U.S. energy series were used for technical assessment; organizational work and supply delays were simulated.\par\smallskip
\noindent\textbf{Findings} -- R1 improved integration and reporting accuracy in controlled SQL Server tests and added stock-movement records, alerts and recovery controls. On comparable feeds, integration defects fell from 2.90--3.46\,\% in the reconstructed R0 to 0.14--0.49\,\% in R1. For national weekly demand, selected forecasts did not significantly outperform the previous-year mean. Process and disruption simulations suggested benefits for manual work and replenishment delay, but these estimates varied with modelling assumptions and are not field-observed resilience outcomes.\par\smallskip
\noindent\textbf{Research limitations/implications} -- R0 combines original T-SQL with reconstructed tables and emulated client rules; it is not a verified historical deployment. Operational benefits depend on an author-specified manual comparator and require field testing.\par\smallskip
\noindent\textbf{Practical implications} -- Transaction traceability, early exception detection and recoverable records can support decisions when movement data are entered reliably and acted upon promptly.\par\smallskip
\noindent\textbf{Originality/value} -- The study evaluates a source-derived legacy design, implements a revised transactional architecture, and distinguishes experimentally observed information-system behaviour from model-predicted organizational effects.\par\smallskip
\end{abstract}
\par
\noindent\textbf{Paper type} -- Research paper\par

\section{Introduction}\label{sec:1}
\begin{figure}[!htbp]
\centering
\includegraphics[width=\linewidth]{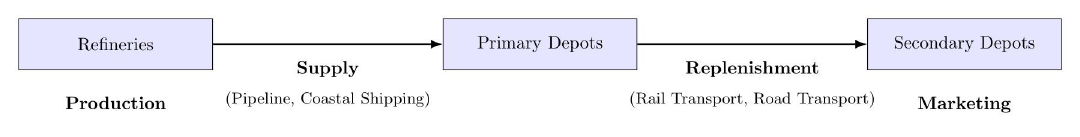}
\caption*{Figure 1. Fuel logistics process: production at refineries, supply of primary depots by pipeline or coastal shipping, replenishment of secondary depots by rail or road, and marketing deliveries to service stations and customers.\par\smallskip 
{\footnotesize
Source: Reproduced from the Author's Original Manuscript
of Loutis and A\"imeur (2026), Figure 1.
https://doi.org/10.1201/9781003647300-28
}}\phantomsection\label{fig:1}
\end{figure}

Fuel distribution depends on timely knowledge of refinery production, depot stocks, replenishment orders, receipts and outbound deliveries. Delays or errors in these records can impair decisions before a physical shortage occurs. Research associates digital integration and information visibility with supply-chain resilience (\citealp{ponomarov2009understanding}; \citealp{brandonjones2014contingent}; \citealp{jiang2024antecedent}), while petroleum monitoring studies mainly address equipment and control systems at individual sites (\citealp{rashad2022smart}; \citealp{fang2025enhancing}; \citealp{waqas2024smart}). This study focuses instead on the integrity and timeliness of transactional information across a fuel distribution network (\hyperref[fig:1]{Figure 1}).

In a conference chapter published in 2026, we described the architecture of a fuel logistics information system and reported simulation-based outcomes (\citealp{loutis2026design}). That account discussed IoT, PLC/FPGA, cloud integration and real-time capabilities. In this follow-up, we distinguish the chapter's architectural scope from functions established by the released Delphi/SQL material. We do not present the earlier simulated outcomes as field measurements or claim an independently verified operational deployment of those integrations.

This paper is an \emph{author-led re-evaluation and extension}, not an independent third-party audit. It compares the available database logic, reconstructed as R0, with a newly implemented R1 artefact under specified test conditions. We ask: RQ1, which information capabilities can be established from the released code? RQ2, how do a revised design and its controls perform in the tested scenarios? RQ3, what contribution to resilience can be inferred from observed information-system behaviour and from explicitly modelled operational processes?

The contributions are a source-level characterization of the released logic, a revised transactional architecture grounded in five design principles, and twelve technical and model-based evaluations. We separate comparable database operations from new R1 capabilities, and measured system outcomes from simulated effects on manual work and replenishment. Detailed defect cases and scenario protocols appear in the appendices.

\hyperref[sec:2]{Section 2} reviews related work and \hyperref[sec:3]{Section 3} the research design. \hyperref[sec:4]{Section 4} presents the artefact and the audit of its implementation, \hyperref[sec:5]{Section 5} the corrections and extensions, \hyperref[sec:6]{Section 6} the evaluation and \hyperref[sec:7]{Section 7} the discussion; \hyperref[sec:8]{Section 8} concludes. Detailed protocols, data dictionaries and additional results are provided in the appendices at the end of this document.

\section{Literature review}\label{sec:2}
\subsection{Digital monitoring in petroleum logistics}\label{sec:2.1}
Digital monitoring of petroleum facilities typically emphasizes PLC/SCADA, IoT sensing and fault detection at terminals or pipelines (\citealp{rashad2022smart}; \citealp{fang2025enhancing}; \citealp{waqas2024smart}; \citealp{ahmed2023fpga}). Related work has explored terminal scheduling and digital twins (\citealp{goncalves2021scheduling}; \citealp{mazzuto2025leveraging}). These contributions complement, but do not directly evaluate, the consistency and latency of fuel-movement records across depots.

\subsection{Digital technologies, information integration and supply chain resilience}\label{sec:2.2}
Visibility, collaboration and information integration are commonly associated with resilience (\citealp{christopher2004building}; \citealp{brandonjones2014contingent}; \citealp{jiang2024antecedent}; \citealp{wu2025digital}). Much of this evidence comes from organizational surveys and reviews (\citealp{iftikhar2024digital}; \citealp{hosseini2025supply}; \citealp{wei2026digital}); it does not itself verify the behaviour of a particular logistics database. Our evaluation links these concepts to traceable movements, reporting, alerting and recovery.

\subsection{Inventory information, forecasting and responsiveness in fuel distribution}\label{sec:2.3}
Inventory control depends on stock position and replenishment policies (\citealp{chopra2016supply}; \citealp{silver2017inventory}). In fuel distribution, responsiveness has been linked to forecast quality (\citealp{mardiana2023gasoline}) and to organizational coordination (\citealp{manggiasih2023evaluating}). The present work evaluates forecast accuracy on held-out public series, separately from the effectiveness of information flows in simulated replenishment. MAPE is accompanied by scaled-error metrics because percentage errors can mislead near zero (\citealp{hyndman2006another}).

\subsection{Modelling methods}\label{sec:2.4m}
Merise specifies conceptual and relational views of information systems (\citealp{avison1991merise}); it may be combined with object-oriented design (\citealp{arab2018unifying}). Here, entity relationships and business constraints define the model under test, with the detailed data dictionary and logical tables provided in Supplementary Section S1.

\subsection{Research gap}\label{sec:2.5}
The research gap is the controlled evaluation of the \emph{transactional information layer} of a multi-site fuel distribution network, rather than a general claim that centralised digitalisation produces resilience. SCADA/PLC primarily concerns physical process monitoring and control, whereas enterprise resource planning (ERP) and supply-chain systems may already provide purchasing, inventory, workflow and reporting functions. The proposed R1 artefact is not claimed to replace or experimentally outperform those systems. Its narrower, testable contribution is the coupling of provenance-aware stock events, database-level validation, exception notifications and operator-reviewed replenishment suggestions with explicit evidence boundaries. \hyperref[tab:I]{Table I} positions the research relative to prior studies; \hyperref[tab:Ia]{Table Ia} distinguishes functional scope without presenting a commercial-system benchmark.

{\footnotesize
\setlength{\tabcolsep}{3pt}
\renewcommand{\arraystretch}{1.15}
\begin{longtable}{@{}L{0.15}L{0.21}L{0.22}L{0.21}L{0.21}@{}}
\caption*{Table I. Positioning relative to prior work}\phantomsection\label{tab:I}\\
\toprule
\textbf{Strand} & \textbf{Representative studies} & \textbf{Focus and evidence} & \textbf{Boundary} & \textbf{Contribution of this study} \\
\midrule
\endfirsthead
\multicolumn{5}{@{}l}{\emph{Table I (continued)}}\\
\toprule
\textbf{Strand} & \textbf{Representative studies} & \textbf{Focus and evidence} & \textbf{Boundary} & \textbf{Contribution of this study} \\
\midrule
\endhead
\bottomrule
\endlastfoot
Control-level monitoring & \citet{rashad2022smart}; \citet{fang2025enhancing}; \citet{waqas2024smart}; \citet{ahmed2023fpga}; \citet{mazzuto2025leveraging} & PLC/SCADA, IoT, FPGA, digital twin; prototypes and laboratory plants & One terminal, pipeline or plant; physical signals & Network-level transactional records; latency and data quality measured \\
Digital technologies, information integration and resilience & \citet{jiang2024antecedent}; \citet{wu2025digital}; \citet{abourokbah2026supply}; \citet{iftikhar2024digital} & Firm-level capabilities; surveys and reviews & No artefact; no test of information delivered & Mechanisms built into an artefact and tested \\
Fuel supply chain analysis & \citet{mardiana2023gasoline}; \citet{manggiasih2023evaluating}; \citet{terab2023information} & System dynamics; soft systems; survey of IT use & Policies and actors; no information system evaluated & Forecasting and information delays evaluated on real and synthetic data \\
Logistics information systems & \citet{biswas2004object}; \citet{trebuna2024logistic}; \citet{loutis2026design} & Object models; logistics IS design; artefact description & Design, prototype demonstration or deployment; no evaluation of the information delivered & Audit, correction and experimental evaluation \\
\end{longtable}
}
\tsrc

{\footnotesize
\setlength{\tabcolsep}{4pt}
\renewcommand{\arraystretch}{1.13}
\begin{longtable}{@{}L{0.19}L{0.26}L{0.30}L{0.25}@{}}
\caption*{Table Ia. Functional positioning; no head-to-head ERP/SCADA benchmark}\phantomsection\label{tab:Ia}\\
\toprule
\textbf{Approach} & \textbf{Primary scope} & \textbf{Relation to the present work} & \textbf{Evidence boundary} \\
\midrule
\endfirsthead
\multicolumn{4}{@{}l}{\emph{Table Ia (continued)}}\\
\toprule
\textbf{Approach} & \textbf{Primary scope} & \textbf{Relation to the present work} & \textbf{Evidence boundary} \\
\midrule
\endhead
\bottomrule
\endlastfoot
PLC/SCADA-oriented monitoring & Field signals and equipment control & Complementary to transactional stock and order records; field integration not evaluated here & Literature-level positioning only \\
ERP and supply-chain suites & Enterprise transactions, inventory and workflows; feature sets vary by product & Potential functional overlap; no claim that R1 implements capabilities unavailable in ERP systems & No commercial product inspected or benchmarked \\
Reconstructed R0 & Legacy Delphi/T-SQL records and reports & Source-derived reference for the shared database operations & Reconstructed schema and emulated forms, not a historical installation \\
Revised R1 & Movement ledger, validation, notifications, access control and order suggestions & Implemented and assessed under specified database tests and process models & No industrial deployment or user study \\
\end{longtable}
}
{\footnotesize\noindent Source: Authors' functional classification and the evidence defined in this study; this table is not an independent comparison of commercial products.\par}\medskip

\section{Research design}\label{sec:3}
\subsection{Design science research approach}\label{sec:3.1}
The study follows design science research (DSR): diagnose an existing design, define requirements, implement a revised artefact and evaluate its behaviour (\citealp{hevner2004design}; \citealp{peffers2007design}). The evaluation is controlled and artificial in the FEDS classification (\citealp{venable2016feds}). An organizational deployment or operator trial has not been performed; the latter is necessary to test the proposed operational benefits.

Three conditions are used. \emph{R0} is a \emph{source-derived comparator}: the released stored procedures, triggers and views are executed as written, but the tables are reconstructed from the available schema descriptions and the missing Delphi form resources require emulation of client checks. Thus R0 is not the original deployed installation, and R0--R1 differences cannot be interpreted as verified improvements over a historical deployment. For two recovery and load tests a separately identified numbering-controlled R0 arm permits continued operation. \emph{R1} is the revised SQL Server implementation developed for this evaluation. \emph{MANUAL} is an author-specified reference workflow used only in the process and replenishment simulations. SQL Server 2022 Developer (16.0.4325.4) ran on Linux with a 2\,GB memory limit and twelve logical CPUs; testing used synthetic transactions generated and submitted by Python programs.

Tests distinguish \emph{inspected} code behaviour, \emph{measured} SQL Server or public-series outcomes, \emph{modelled} process and discrete-event estimates, and functions \emph{not assessed} without users or field data. The principal R1 results were evaluated with synthetic seeds 101--105 where stated, scaled networks and an out-of-sample forecasting interval; T8 one-at-a-time sensitivity checks used only seeds 101--103. The evaluation protocols and numerical summaries are presented in the article and appendices.

\subsection{Requirements and design principles}\label{sec:3.2}
Resilience is treated as the capability to prepare for, respond to and recover from disruptions (\citealp{ponomarov2009understanding}). Following the distinction between resource-oriented and action-oriented digital technologies (\citealp{wu2025digital}), the study assumes that an information system contributes to these phases through the information it provides and the actions it enables: visibility of stock and flows, an ingredient of supply chain agility (\citealp{christopher2004building}) that has been found to support resilience (\citealp{brandonjones2014contingent}), supports readiness; in the study's reasoning, early and correct signals support response, and recoverable records support recovery of the information function itself. Seven requirements follow from these phases and from the capabilities claimed for the system: REQ1 complete and consistent records of every stock movement; REQ2 timely visibility of stock and stock in transit; REQ3 correct reports generated without manual consolidation; REQ4 anticipation of demand and depot autonomy; REQ5 early detection of critical stocks, anomalies and late deliveries; REQ6 few manual interventions and few remaining errors; REQ7 controlled access, recoverability and stable performance as the network grows. \hyperref[tab:II]{Table II} states five design principles that answer them, each with its resilience mechanism, its implementation in R1 and its test.

{\footnotesize
\setlength{\tabcolsep}{3pt}
\renewcommand{\arraystretch}{1.15}
\begin{longtable}{@{}L{0.20}L{0.24}L{0.22}L{0.18}L{0.16}@{}}
\caption*{Table II. Design principles of the revised artefact}\phantomsection\label{tab:II}\\
\toprule
\textbf{Design principle (requirements)} & \textbf{Mechanism} & \textbf{Resilience rationale} & \textbf{Implementation (Table VII)} & \textbf{Tests} \\
\midrule
\endfirsthead
\multicolumn{5}{@{}l}{\emph{Table II (continued)}}\\
\toprule
\textbf{Design principle (requirements)} & \textbf{Mechanism} & \textbf{Resilience rationale} & \textbf{Implementation (Table VII)} & \textbf{Tests} \\
\midrule
\endhead
\bottomrule
\endlastfoot
DP1 One ledger of physical movements (REQ1--REQ3) & Every stock change is recorded once with its source document; balances, occupancy and reports are derived from it & Readiness: visibility of stock and flows (\citealp{christopher2004building}; \citealp{brandonjones2014contingent}) and traceability of each change to its document & C2, C3, E1, E2, E3 & T1, T3, T12 \\
DP2 Validate at entry, reconcile at the count (REQ1, REQ6) & Invalid and contradictory entries are rejected; physical events are always recorded and, when inconsistent, flagged; counts are compared with the book & Decisions rest on data that can be trusted; losses and wrong records become visible & C4, C5, E3, E6 & T1, T7, T12 \\
DP3 Push exceptions as events (REQ2, REQ5) & Alerts on the inventory position, abnormal movements, discrepancies and overdue deliveries, notified when the triggering record is committed & Response: shorter information delay between a condition and the decision maker & E4, E5 & T2, T5, T8 \\
DP4 Propose what stored records determine (REQ6) & Orders pre-filled from routes, policy parameters and the inventory position; the operator reviews and confirms & Response: shorter decision delay and fewer manual interventions, without automatic commitment & E8 & T6, T8 \\
DP5 Separate write rights and keep a recoverable history (REQ7) & Writes only through operations granted per role; hashed credentials with lockout; full and frequent transaction-log backups & Recovery: continuity of the information function after failures and misuse & E7 & T9, T11, T12 \\
\end{longtable}
}
\tsrc

The five principles operate as a coupled, context-specific design mechanism rather than five independent software features. DP1 and DP2 link a physical movement to a unique source record and distinguish an issued order from a completed receipt; an apparently timely balance is not reliable if that distinction is lost. DP3 and DP4 connect an alert to an operator-reviewed action, but their effectiveness is conditional on when the underlying physical event is entered and when a responsible operator can respond. DP5 preserves the integrity and recoverability of the informational state; it does not restore disrupted physical transport. These constraints yield testable predictions about record correctness (T1, T3), record-to-display delay (T2, T5), and event-to-order delay (T8), while setting a boundary on claims about organizational resilience (T12).

\subsection{Evaluation framework}\label{sec:3.3}
\hyperref[tab:III]{Table III} defines the twelve scenarios, comparators and measures. Only shared functionality is compared directly between R0 and R1; new R1 capabilities are evaluated as additional coverage. Technical counts and latencies come from controlled software tests. Human actions, entry errors, decision times and disruption outcomes are model estimates relative to the specified MANUAL workflow. The appendices contain scenario definitions, parameters and more detailed outputs.

{\footnotesize
\setlength{\tabcolsep}{3pt}
\renewcommand{\arraystretch}{1.15}
\begin{longtable}{@{}L{0.13}L{0.27}L{0.15}L{0.25}L{0.20}@{}}
\caption*{Table III. Evaluation scenarios, units, arms and measures}\phantomsection\label{tab:III}\\
\toprule
\textbf{Scenario} & \textbf{Unit and procedure} & \textbf{Arms} & \textbf{Measure} & \textbf{Reported as} \\
\midrule
\endfirsthead
\multicolumn{5}{@{}l}{\emph{Table III (continued)}}\\
\toprule
\textbf{Scenario} & \textbf{Unit and procedure} & \textbf{Arms} & \textbf{Measure} & \textbf{Reported as} \\
\midrule
\endhead
\bottomrule
\endlastfoot
T1 Data integration & Feed row; six feeds of 14 days with seeded defects of ten types; one year of operation (T1b) & R0 (form, bulk import), R1 (direct, staged) & Loss of valid rows, contamination by defective rows, integration defect rate, balance consistency & Rates with numerators and denominators, per seed and defect rate \\
T2 Monitoring & Committed stock change; paced replay with push and polling dashboards & R1 (R0 has no refresh) & Submission-to-display latency & Median, 95th percentile and maximum \\
T3 Reporting & Report field; nine reports against independent reference values & R0 (published, logic only), R1 & Field accuracy; generation time & Correct fields of all fields; maximum time \\
T4 Forecasting & Forecast origin; public weekly data and synthetic depot data, rolling origin & Previous-year aggregate (R0) and nine methods & MAPE, MAE, RMSE, MASE & Test-period error per series and method; paired comparisons on the same forecast origins (block-bootstrap intervals, Diebold--Mariano tests, Holm adjustment) \\
T5 Alerting & Expected alert; same replay as T2 with seeded disruptions & R1 (R0 has no alerts) & Recall, precision, notification latency & Raised, missed and false alerts; median and maximum latency \\
T6 Manual interventions & Business action per operating day; process model & MANUAL, R0, R1 & Actions and time per day & Per day; change relative to MANUAL \\
T7 Manual errors & Seeded-error probe (measured); erroneous value reaching the record (modelled) & MANUAL, R0, R1 & Interception share; remaining errors & Intercepted probes; remaining errors per day with Monte Carlo intervals \\
T8 Supply delay & Replenishment order; discrete-event model & MANUAL, R0, R1 & Position below reorder point to arrival, by component & Mean and components; change relative to MANUAL with range over seeds and sensitivity \\
T9 Security & Access attempt (role $\times$ operation); authentication cases & R0 (forms, database), R1 & Unauthorized operations permitted & Counts of permitted and refused attempts \\
T10 Scalability & Request; 1$\times$, 2$\times$, 3$\times$, 5$\times$ network, history and users & R0 (logic), R1 & 95th-percentile response time, failures & Per load level \\
T11 Backup and recovery & Restoration trial after simulated media failure & R0 strategy, R1 strategy & Restoration time, data lost & Per trial \\
T12 Resilience & Disruption scenario & MANUAL, R0, R1 & Supply delay, stock-out hours, unmet demand, time to detect & Mean over seeds per scenario \\
Usability & Task attempt by an operator & -- & Unaided task completion & Not testable here (no participants) \\
\end{longtable}
}
\tsrc

\section{The artefact and the audit of its implementation}\label{sec:4}
\subsection{Architecture and roles}\label{sec:4.1}
The available system uses a Delphi VCL client and SQL Server through ADO components (\citealp{embarcadero2025rad}; \citealp{microsoft2026sqlserver}); its code includes 46 form units, three stored procedures, two triggers and eleven SQL views. Functions documented in the earlier chapter but not evidenced by the released code are identified separately in \hyperref[tab:V]{Supplementary Table V}. Figure 2 presents the legacy application roles and shared server, while Figure 3 presents a schematic network of refineries and multiple depots sharing logistics transaction records.

\begin{figure}[!htbp]
\centering
\includegraphics[width=\linewidth]{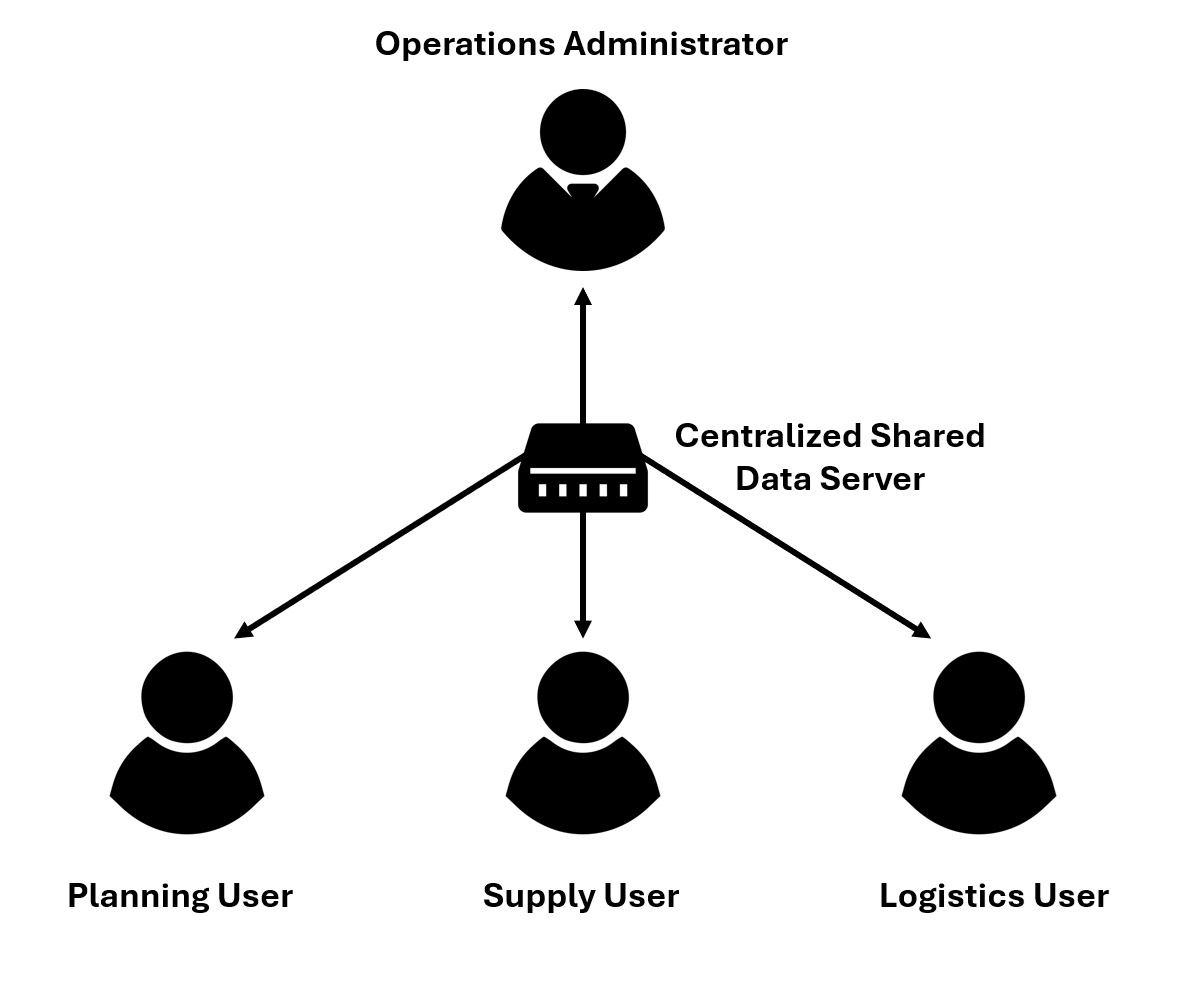}
\caption*{Figure 2. Legacy client--server arrangement with four application roles and a shared database server.\par\smallskip 
{\footnotesize
Source: Adapted from the Author's Original Manuscript
of Loutis and A\"imeur (2026), Figure 2.
https://doi.org/10.1201/9781003647300-28
}}\phantomsection\label{fig:2}
\end{figure}

\begin{figure}[!htbp]
\centering
\includegraphics[width=\linewidth]{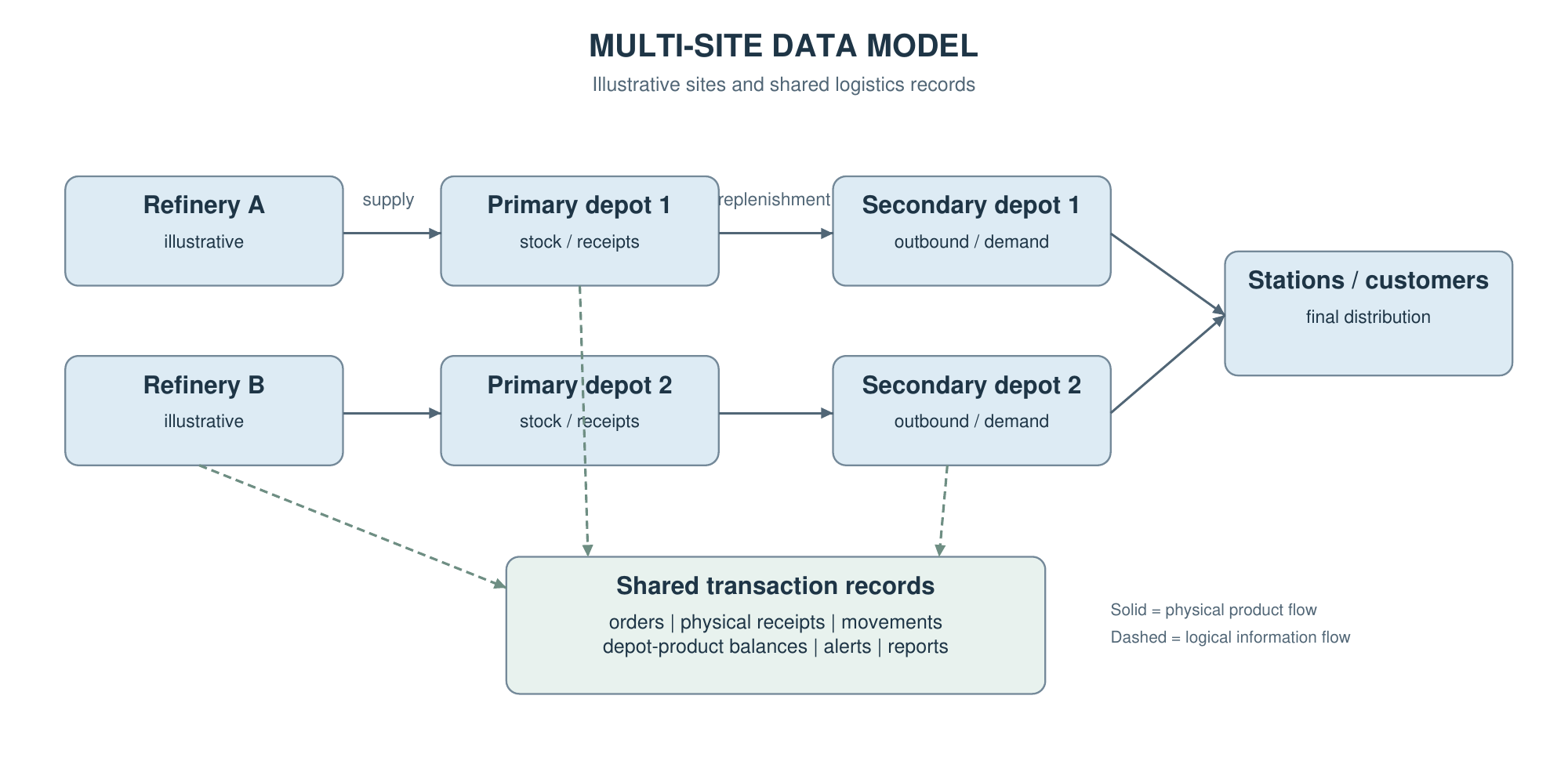}
\caption*{Figure 3. Schematic multi-site fuel distribution network and shared transactional records. Solid arrows indicate possible physical logistics flows; dashed arrows indicate information flows to the shared system.\par\smallskip
{\footnotesize Source: Authors' own schematic, based on the process model in \citet{loutis2026design} and the requirements of this study.}}
\phantomsection\label{fig:3}
\end{figure}

\subsection{Conceptual data model}\label{sec:4.2}
The Merise conceptual data model (CDM) contains thirteen entities and seventeen associations (\hyperref[tab:IV]{Supplementary Table IV}); the data dictionary is in Supplementary Tables S1a--S1b. It describes refineries, depots, products, transporters, production records and supply or replenishment orders. Figure 4 presents an EER-style representation of the legacy relational model derived from the published conceptual descriptions and Delphi/T-SQL declarations. The diagram identifies the distinct emitting and receiving depot roles of replenishment orders. R1 extends this model with separate records for outbound deliveries, physical receipts and stock counts.

\clearpage
\begin{landscape}
\begin{figure}[!htbp]
\centering
\includegraphics[width=0.96\linewidth,height=0.74\textheight,keepaspectratio]{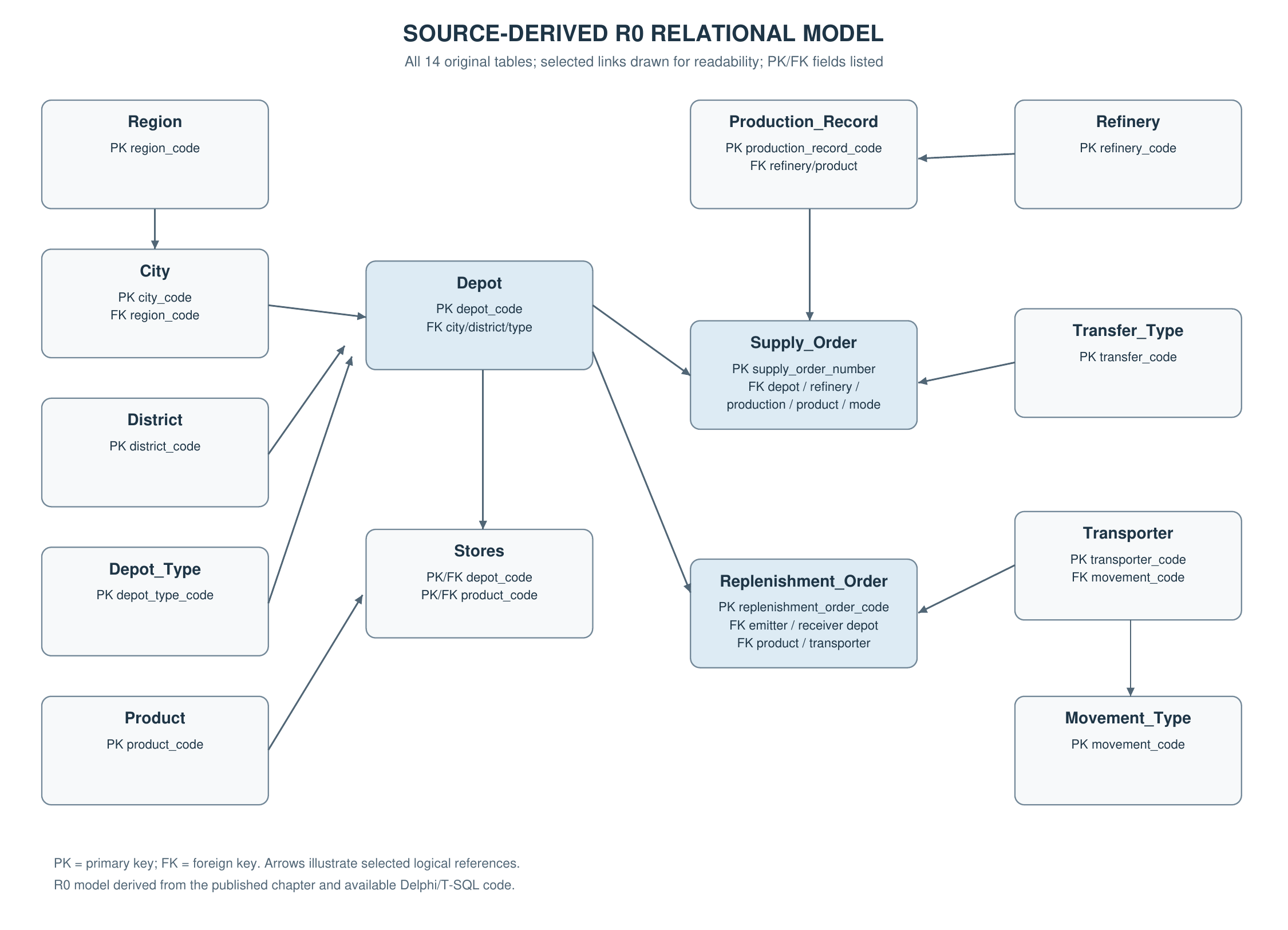}
\caption*{Figure 4. EER-style representation of the source-derived R0 relational model. Lines show selected logical key references and associations; attribute lists are restricted to identifying and linking fields.\par\smallskip
{\footnotesize Source: Authors' reconstruction of the model in \citet{loutis2026design}; adapted data-model concepts, newly drawn visualisation.}}
\phantomsection\label{fig:4}
\end{figure}
\end{landscape}

\subsection{Logical data model and demonstration scenario}\label{sec:4.3}
The CDM maps to fourteen original relational tables (Supplementary Table S1c). R1 extends the model with stock movements, outbound deliveries, receipt states, counts, alerts, supply routes and staging records (Supplementary Table S1d). A worked synthetic transaction and the calculation rules are documented in S1e and S2. Quantities are expressed in cubic metres.

\subsection{Audit of the published implementation}\label{sec:4.4}
The released material was inspected at commit \texttt{ac3d536} (14 November 2026). It provides executable T-SQL and Delphi units but not the complete form resources or installation schema needed to reproduce the graphical application. The available SQL logic can therefore be tested only with a reconstructed data schema, while UI checks must be inferred from source. As authors of the 2026 chapter, we clarify that its descriptions of IoT, PLC/FPGA, cloud, alerts and forecasting were broader than the capabilities evidenced by this released implementation; no present experiment relies on an operational field-device or cloud installation. The comparison below concerns the material available for this study, not an independently verified historical deployment (\hyperref[tab:V]{Supplementary Table V}).

Key observed defects include document numbering based on a fixed-width \texttt{COUNT(*)+1} suffix, triggers that process only one row of a multi-row insert, and a report that cross-joins order tables before aggregation. A one-year synthetic replay exposed numbering rejection, while focused SQL tests reproduced the trigger and reporting failures (\hyperref[tab:VI]{Supplementary Table VI}). These findings characterize the released logic in the reconstructed R0 environment and motivate the controls incorporated into R1.

The published design also has strengths that R1 keeps: codes are selected from lists rather than typed, document numbers and total costs are computed, stocks are updated by the database rather than by hand, the supply form checks the remaining production quantity, and order dates cannot be set before the entry date.

\section{Corrections and extensions: the revised artefact}\label{sec:5}
R1 replaces fragile document numbering, records stock movements in a ledger, separates orders from physical receipts, validates feeds and flags discrepancies. Additional functions comprise stock alerts, update notification, role-based database permissions, backup/recovery controls and operator-reviewed order suggestions (\hyperref[tab:VII]{Supplementary Table VII}). The five design principles in \hyperref[tab:II]{Table II} connect these mechanisms to information visibility, accuracy and response.

R1 was evaluated as a single defined configuration. Database controls, feed staging, an event-driven notification channel and decision-support functions were exercised in twelve scenarios; only shared operations permit like-for-like comparisons with reconstructed R0.

Compatibility-oriented SQL corrections were also exercised on the reconstructed legacy schema in 57 tests, including integrity, reports, regression, migration and backup/restore. This tests database-level corrections, not operation of the original Delphi client; extensions requiring new entities or interfaces were evaluated only as part of R1. Figure 5 depicts the principal R1 data-model extensions. It distinguishes stock movements, outbound deliveries, physical counts, alert rules and alert records as separate entities rather than treating every order as an immediate physical stock change. The relationships are derived from the R1 design specified in Supplementary Table S1d; the diagram is not an independently audited production schema.

\emph{Illustrative operator decision.} Consider the synthetic worked transaction in Supplementary Table S1e: a primary depot receives 6,000~m$^3$ and dispatches 2,500~m$^3$ to a secondary depot. In R1, the order itself does not establish physical receipt; the corresponding recorded movement changes the stock ledger. Inventory position and days of cover can then be calculated from the current recorded state, and a threshold exception can prompt an operator to review a proposed replenishment order. This sequence explains the decision mechanism without claiming a real-world operator trial.

\clearpage
\begin{landscape}
\begin{figure}[!htbp]
\centering
\includegraphics[width=0.96\linewidth,height=0.74\textheight,keepaspectratio]{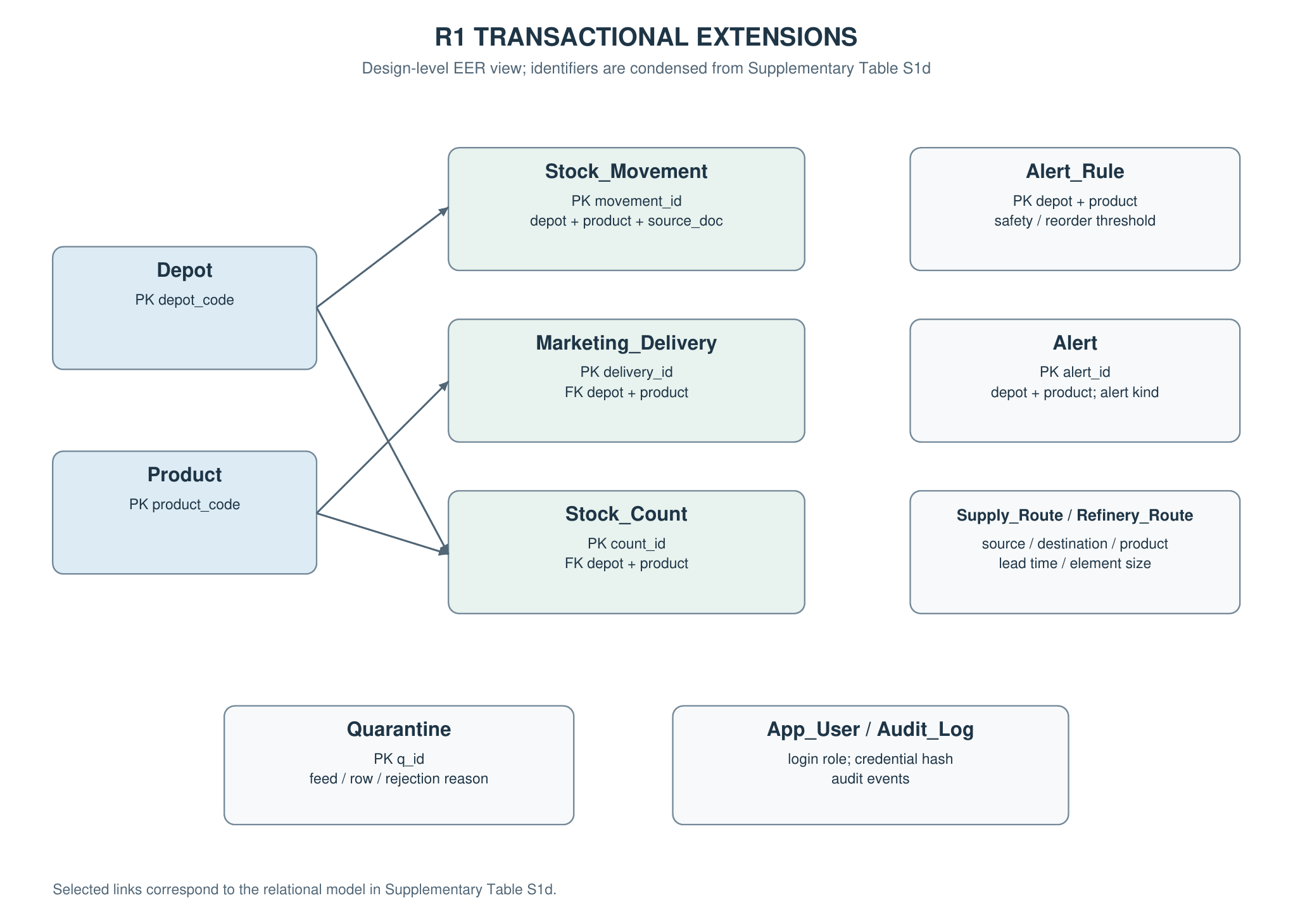}
\caption*{Figure 5. Selected relational extensions of R1, illustrating links between depot-product records, movements, deliveries, counts, alert rules and notifications. Additional columns and constraints are documented in Supplementary Table S1d.\par\smallskip
{\footnotesize Source: Authors' own work.}}
\phantomsection\label{fig:5}
\end{figure}
\end{landscape}

\section{Evaluation results}\label{sec:6}
The following results concern controlled SQL Server tests of the final R1 design, synthetic validation seeds 101--105, and public forecasting observations. Detailed counts and distributions are reported in the supplementary tables. \hyperref[tab:XIV]{Table XIV} summarizes the evidence by method.

\subsection{Data integration and sustained operation (T1)}\label{sec:6.1}
Fourteen-day synthetic feeds contained valid rows and seeded defects. On the four feeds represented by both designs, the integration defect rate in R0 was 2.90--3.46\,\%, versus 0.14--0.49\,\% with R1 staging. The additional outbound and receipt feeds were assessed only for R1. The staged validator retained unresolved valid rows in quarantine; balance consistency was not a like-for-like outcome because the two systems use different posting rules (\hyperref[tab:VIII]{Supplementary Table VIII}).

In one year of nominal synthetic operation, reconstructed R0 accepted 3,009 of 12,351 production records and orders after the document-number suffix overflowed; R1 accepted all 12,351. The result applies to the tested R0 reconstruction and is not an observation of the historical installation.

\subsection{Automated reporting (T3)}\label{sec:6.2}
On the synthetic report inventory, seven of nine assessed legacy report views returned correct fields; the depot status was affected by stock-occupancy arithmetic, and the general supply forecast failed with an SQL integer overflow. For the eight shared report layouts, R0 produced 4,697/4,733 correct fields and R1 produced 4,733/4,733. The ninth report, a general supply forecast, has different layouts: 128 reference fields in R0 (the query failed) and 50 in R1 (all correct). Thus R1 also attained 4,783/4,783 fields across its own nine-report inventory (4,733 shared-layout fields plus 50 forecast fields); R0's corresponding inventory had 4,861 fields (4,733 plus 128). The two total inventories are not field-for-field comparable. R1 generated its reports in at most 37 ms in the nominal test (Supplementary Table S3c).

\subsection{Monitoring and alerting (T2, T5)}\label{sec:6.3}
R1 added stock-update notification and alert functions absent from the examined R0 code. These functions were assessed independently of the R0 comparison (\hyperref[tab:IX]{Supplementary Table IX}).

In the controlled replay, R1 raised all 244 expected alerts at nominal load and all 1,144 at fivefold load, with no false positives in the seeded cases. Median push-display latency remained below 0.01 s, but fivefold-load maximum delays reached 7.327--9.061 s during the early part of the test window; the full-window push P95 was 0.806--1.306 s across runs. These spikes remain unexplained and must not be omitted from responsiveness claims. Additional notification and polling results are in Supplementary Table IX. The tested readout is a database-backed monitoring client rather than the original Delphi GUI: its operational indicators comprise depot-product stock position, stock in transit, days of cover (stock divided by the preceding 28-day mean outbound quantity) and open alert count (Supplementary Section S2.2). The controlled replay verifies record visibility and notification latency; it does not validate dashboard usability, operator interpretation or a field installation. No screenshot of an independently tested graphical interface is provided, and none of the schematic figures should be described as one.

\subsection{Stock and demand forecasting (T4)}\label{sec:6.4}
Rolling-origin forecasting used public EIA series, with 2021--2023 for method selection and 2024--2026 for testing (\citealp{hyndman2021forecasting}). M0 is the previous-year mean and M1 a one-step naive comparator; the candidate set also includes Holt--Winters (M4), harmonic regression with ARIMA(1,1,1) errors (M5), and a variant of Holt--Winters with the 2020 pandemic observations replaced by the corresponding 2019 observations (M9), as specified in Supplementary Section S3 and Table X. The comparative results appear in Supplementary Table X. The interpretation does not infer a significant gain in national weekly demand forecasting from differences in raw MAPE alone.

Selected methods did not establish an improvement over the previous-year mean for any of the three national weekly demand series after multiple-testing adjustment. They were more accurate on gasoline and distillate stock series. One-week test MAPEs for selected methods were 2.96\,\%, 5.99\,\% and 6.84\,\% on the three demand series, and 0.83\,\% and 1.59\,\% on the two stock series. Paired forecast comparisons, uncertainty intervals and depot-level synthetic results are in Supplementary Section S3.

\subsection{Security, scalability and recovery (T9, T10, T11)}\label{sec:6.5}
\emph{Access control.} Of the 25 role--operation pairs that the role policy forbids and that the published model can express, seven were reachable from the menus of the corresponding role (for example, the logistics role can open the supply-order form and edit master data), and all 25 succeeded when executed through the single application login that the client code implies. Passwords were stored in clear text, the role was the user name, and 1,000 wrong guesses were rejected without any lockout, after which the correct password was still accepted. In R1, none of 36 forbidden attempts succeeded (none of the 25 that the published model can also express) and all 32 permitted ones did; passwords were stored as salted PBKDF2 hashes (210,000 iterations, an implementation parameter rather than a requirement of NIST guidance; \citealp{temoshok2025digital}), an account was locked after five failures for 15 minutes, and database logins with wrong or empty passwords were rejected. No forbidden operation succeeded in the tested matrix. This is not a penetration test, transport encryption was not tested, and neither version underwent an information security risk assessment following ISO/IEC 27005 (\citealp{isoiec27005_2022}) or a review against the control catalogue of NIST SP 800-53 (\citealp{nist2020security}); R1's separation of database roles, operation-level grants and hashed credentials with lockout is an instance of the layered protection that the catalogue calls defense in depth (control PL-8(1)), not an assessed implementation of it.

\emph{Scalability.} \hyperref[tab:XI]{Supplementary Table XI} gives the response times per load level (additional load levels summarized in Supplementary Table S3d). Both versions were measured on SQL Server on the same date, with the same calendar of 686 days. With the published logic, the general supply forecast failed in every request at every level (33 of 33 at the nominal load, 99 of 99 at 5$\times$), always with the arithmetic overflow of \hyperref[sec:6.2]{Section 6.2}, after up to 3.6 s at the nominal load and up to 54.5 s at 3$\times$ in T10's concurrent mixed-request workload, which is not directly comparable to T3's isolated report timing; the highest 95th percentile of the other operations was 1.309--1.359 s (the previous-year reports at 5$\times$). In R1, the slowest operation at 5$\times$ was the general supply forecast, with a 95th-percentile response time of 0.145--0.178 s across three runs; the tabulated run had a 99th percentile of 0.181 s and a maximum of 0.200 s. Over all 3,509 requests at 5$\times$, the mean was 0.039 s, the median 0.047 s, the 95th percentile 0.095 s, the 99th percentile 0.128 s and the maximum 0.285 s, and no request failed; at the nominal load, R1's slowest operation (transaction entry) had a 95th percentile of 0.050 s. The request mixes have the same shares but not the same content for one class: R1's stock read is its dashboard, which also counts open alerts and stock in transit, whereas R0 reads the stock table; reports, transaction entry and the general supply forecast are common operations. Without R1's indexes, the 95th percentile of the slowest operation at 5$\times$ was 0.160 s (the general supply forecast; 3,494 requests, no failure): at this size, the indexes changed little, and the forecast's gain comes from removing the cross join.

\emph{Recovery.} Under three simulated media failures, full and 60-s transaction-log backups in R1 reduced the recovery-point interval compared with a once-daily full-backup strategy for the reconstructed R0. R1 restored in 0.84--0.97 s and lost 58--106 recently committed rows; the full-backup strategy lost all 40 documents entered since its backup. These small-database tests establish recovery-point behaviour, not operational recovery at scale (Supplementary Table S3g).

\subsection{Manual interventions and errors (T6, T7)}\label{sec:6.6}
\emph{Seeded errors (measured).} R1 intercepted 24 of 30 designed error probes, versus 7 of 30 in the reconstructed R0; on the 27 error types supported by both, the counts were 21 versus 7. Neither detected erroneous values consistent with all other records (\hyperref[tab:XII]{Supplementary Table XII}).

\emph{Process outcomes (modelled).} Under the specified synthetic workload and author-defined MANUAL workflow, the estimated business-action count was 1,670.6 per day manually versus 526.0 with R1 (68.5\,\% fewer). Modelled working-time and error reductions also depended on assumed typing, checking and error rates. These figures are simulation outputs, not reductions observed in an organization; the full task decomposition and uncertainty ranges appear in Supplementary Tables S3a and S3h.

\subsection{Supply delay and operational resilience (T8, T12)}\label{sec:6.7}
\hyperref[tab:XIII]{Supplementary Table XIII} decomposes the supply delay. In the manual procedure, a depot-product whose inventory position fell below its reorder point received its replenishment on average 73.7 h later: 24.0 h before the information reached the decision maker, 6.0 h to prepare, approve and transmit the order, 10.8 h waiting for a loading slot and 33.0 h in transit. The published system shortened this to 63.6 h (13.7\,\% less), mainly by faster order entry; information still arrived through the next morning's counts. R1 with outbound deliveries posted at 20:00 reached 55.3 h (25.0\,\% less): the alert follows the posting immediately, but an evening alert waits for the next working morning, which moves part of the delay from information to decision. Posting deliveries also at 11:00, the second iteration, gave 46.5 h, a reduction of 36.9\,\% (36.0--38.3\,\% over five seeds), and reduced the order delay from the reorder point to the issued order by 56.1\,\% (29.9 to 13.1 h). An hourly meter feed, which is not implemented, would reach 44.3 h (39.8\,\%). Transit, which no information system shortens, is 45\,\% of the manual delay: removing every information, decision and dispatch delay would leave transit alone and bound the reduction at about 55\,\%, a bound that only continuous recording of outbound flows and around-the-clock response could approach. A one-at-a-time sensitivity analysis on three of the five validation seeds (101--103) gave reductions of 30.6--40.3\,\% with two postings across the stated preparation, approval, response and report-time assumptions; assuming the manual morning report missed the daily review raised that estimate to 75.8\,\%. Moving the single evening posting to 16:00 or 22:00 gave 15.7\,\% or 26.9\,\%, respectively, versus 25.0\,\% at 20:00. These sensitivity bounds describe only the three tested seeds and must not be attributed to all five validation seeds. With the same reorder points, the modelled manual procedure produced 423.6 stock-out depot-product hours over the entire 150-day horizon across 96 depot-products (fill rate 99.82\,\%), compared with 7.6 for reconstructed R0 and none for R1. The 150-day totals include the simulation warm-up period; they are not post-warm-up steady-state estimates.

Across specified demand and transport disruptions, the discrete-event model predicted fewer stock-outs and less unmet demand with R1 than with the author-specified manual process, under the chosen inventory, reporting and decision policies. For example, a 60\,\% demand surge raised mean manual stock-out time to 502.6 h in the simulated network while R1 recorded none; such outcomes are conditional on these model settings, not measured organizational resilience. Full scenarios appear in Supplementary Table S3h.

\subsection{Summary of results}\label{sec:6.8}
\hyperref[tab:XIV]{Table XIV} separates measured information-system results from modelled operational estimates. The revised system improved record completeness and the tested controls; the extent to which this yields faster replenishment in practice depends on reporting frequency, human response and physical transit. Public-series demand forecast accuracy did not significantly exceed the previous-year-mean benchmark.

{\small
\setlength{\tabcolsep}{4pt}
\renewcommand{\arraystretch}{1.12}
\begin{longtable}{@{}L{0.30}L{0.48}L{0.22}@{}}
\caption*{Table XIV. Core findings and evidential scope}\phantomsection\label{tab:XIV}\\
\toprule
\textbf{Evaluation} & \textbf{Core finding} & \textbf{Basis} \\
\midrule
\endfirsthead
\multicolumn{3}{@{}l}{\emph{Table XIV (continued)}}\\
\toprule
\textbf{Evaluation} & \textbf{Core finding} & \textbf{Basis} \\
\midrule
\endhead
\bottomrule
\endlastfoot
Data integration (T1) & Shared-feed integration defect rate: 2.90--3.46\,\% (R0) versus 0.14--0.49\,\% (R1) & Measured; synthetic, reconstructed comparator \\
Reporting (T3) & R1: 4,733/4,733 fields in eight common layouts; 4,783/4,783 in its full nine-report inventory & Measured; synthetic; layouts differ \\
Monitoring and alerts (T2, T5) & All seeded alerts raised; push maxima up to 9.061 s at fivefold load & Measured; synthetic \\
Forecasting (T4) & No established improvement over previous-year mean for national weekly demand & Public-series tests \\
Manual work and errors (T6, T7) & Fewer actions and residual errors predicted under the study's process assumptions & Modelled; author-defined MANUAL \\
Supply delay (T8, T12) & Directional model benefit; magnitude strongly sensitive to reporting schedules & Modelled; scenario-dependent \\
Security, load and recovery (T9--T11) & Tested R1 controls, response times and log-backup recovery & Measured; development environment \\
\end{longtable}
}
{\footnotesize\noindent Note: For T3, 4,733 fields belong to the eight directly comparable layouts. The ninth report has 128 fields in R0 and 50 in R1, so the total inventories are 4,861 and 4,783 fields, respectively; total-inventory percentages are not direct field-level comparisons. Source: Authors' own work\par}\medskip

\section{Discussion}\label{sec:7}
\subsection{Answers to the research questions}\label{sec:7.1}
\emph{RQ1.} The author-led source review identified a relational core and several implementation weaknesses, including numbering, trigger and reporting defects. Some capabilities described in the earlier chapter could not be established from the released implementation. This finding concerns source evidence and is not an independently verified assessment of its historical installation.

\emph{RQ2.} R1 improved data validation and report correctness and provided separately tested notification, security and recovery functions (\hyperref[tab:XIV]{Table XIV}). Measured results apply to controlled SQL Server experiments and to the reconstructed R0 comparator. Organizational effects remain model estimates under defined scenarios.

\emph{RQ3.} The design can support readiness through reliable stock records, response through timely notifications and decision support, and information recovery through retained logs. These are information-system mechanisms, not direct evidence that a fuel supply chain recovered from an actual disruption.

\subsection{Contribution}\label{sec:7.2}
In the design-science contribution framework, this is an \emph{improvement} in the sense of \citet{gregor2013positioning}: it does not propose a new general theory of resilience. Its theoretical contribution is a conditional account of how transactional controls may translate into readiness and response at network level (\citealp{jiang2024antecedent}; \citealp{wu2025digital}), beyond equipment-level monitoring (\citealp{rashad2022smart}; \citealp{waqas2024smart}). The principles identify a chain of mechanisms with identifiable failure boundaries, rather than assuming that digitalization itself causes resilience.

\emph{P1, event provenance and valid state (DP1--DP2):} when receipts, deliveries and corrections are recorded as distinct events linked to source documents, ledger-derived balances and reports should exhibit fewer untraceable discrepancies than overwritten stock totals; this prediction depends on complete physical-event entry and can be checked by reconciled depot-product balances (T1, T3). \emph{P2, recording cadence and actionable signals (DP3--DP4):} rapid notifications should shorten record-to-display delay, but will shorten event-to-order delay only when physical events are entered before the relevant review window and an operator responds; T2/T5 test the first link technically, while T8 tests the conditional second link only in a model. \emph{P3, recoverable information versus physical constraints (DP5):} access controls and log backups can maintain or restore usable records, but cannot remove dispatch and transit time; T9/T11 assess information controls, whereas T12 models the operational consequence. These propositions are analytically derived and partly illustrated, not established as field-level causal effects.

\subsection{Size and limits of the effects}\label{sec:7.3}
\emph{Supply delay.} The estimate is particularly sensitive to reporting schedules. The model predicted a 25.0\,\% reduction with outbound deliveries posted once daily and 36.9\,\% when posted twice daily, relative to the author-defined manual workflow. One-at-a-time schedule and parameter checks on three validation seeds, rather than the five used for the principal result, yielded estimates from 15.7\,\% to 75.8\,\% across different assumptions and scenarios. These are conditional sensitivity results, not a five-seed uncertainty interval or an observed operational benefit. The reported stock-out totals also include the model warm-up period.

\emph{Forecasting.} Selected methods performed well on two stock series, but their improvements for national weekly demand were not established against the previous-year mean after adjustment. Aggregation and series variability materially affected error levels; the full paired comparisons appear in Supplementary Table S3i.

\emph{Manual work and errors.} Modelled reductions reflect assumed entry rates, checking probabilities and process design (\citealp{wahi2008reducing}; \citealp{panko2000spreadsheet}). They must be validated with operators; the technical error probes do not cover plausible but internally consistent mistakes.

\subsection{Implications for practice}\label{sec:7.4}
Three practical lessons follow. First, sustained-volume tests and set-based transaction checks expose defects that small demonstrations may miss. Second, a coherent movement ledger enables traceable stocks and reports, provided physical events are entered. Third, notification performance alone is insufficient: review schedules, responsibility for open alerts and physical transit determine when decisions can change outcomes. For example, the T8 discrete-event experiment gave a modelled mean replenishment interval of 73.7~h under MANUAL and 46.5~h under R1 with two daily postings; the difference is conditional on the specified reporting and decision timetable, not evidence of shorter physical transport or a deployment benefit. In practical terms, the system is intended to make a recorded shortage and the corresponding source documents visible before the next decision window, rather than to remove the transport constraint.

\subsection{Limitations and future research}\label{sec:7.5}
R0 uses original SQL on reconstructed tables and inferred Delphi checks; it does not reproduce an authenticated historical deployment. Technical tests were performed on SQL Server 2022 Developer under synthetic loads on a memory-limited workstation. The author-defined MANUAL workflow, simulated recording delays and error probabilities underpin the estimated business actions, error reductions and supply delays. A field or operator study has not been conducted. Public weekly U.S. data support only the forecasting comparisons, not the operational resilience claims.

Push-display delays peaked at 9.061 s during the early load window. Future work should evaluate user interaction and performance in operational settings, and extend the system to automatic measurements and multi-product shipments.

\section{Conclusion}\label{sec:8}
This author-led re-evaluation and extension of an earlier fuel logistics information system presents a revised transaction architecture and twelve evaluations. Controlled SQL Server experiments on synthetic data found improved integrity, correct reports and functioning alerts and recovery controls relative to a \emph{reconstructed} source-derived comparator. The public-series forecasting tests did not establish better accuracy than a previous-year-mean baseline for national demand. Process and disruption models indicate possible benefits for work and replenishment, including estimated reductions in business actions, residual errors and delay, but the magnitudes depend strongly on the defined manual workflow and reporting assumptions. The study therefore supports the value of reliable information systems as potential resilience enablers; whether the predicted organizational benefits occur requires operator and field validation.

\section*{Data, code and supplementary material}

\noindent\textbf{Figure permissions and acknowledgements.} Figures 1 and 2 are from an Author's Original Manuscript
of a book chapter published by Routledge/Taylor \&
Francis in \emph{Moving Towards Sustainable Businesses
and Developing More Resilient Economies} on May 13,
2026, available online at
\url{https://doi.org/10.1201/9781003647300-28}. This licence does not allow the use of the Publisher's
version/PDF (this is the version of record that is
published on the publisher's website) to be posted
online. From: \emph{Moving Towards Sustainable Businesses and Developing More Resilient Economies}, 1st edition,
edited by Aamir Rashid and Fenio Annansingh-Jamieson,
Copyright \textcopyright{} 2026 by CRC Press.
Reproduced by permission of Taylor \& Francis Group.

\textbf{Relation to prior publication.} This article extends the authors' 2026 conference chapter (\citealp{loutis2026design}). The earlier chapter presented a broad architectural design and simulation-based performance statements. The present study reports an author-led re-evaluation of released source material and a separately evaluated revised design; neither the prior simulated outcomes nor a historical field deployment are treated as independently reproduced.

\textbf{Data availability.} The study used synthetic operational inputs and public weekly U.S. Energy Information Administration series (\citealp{eia2026weekly}, downloaded 13 November 2026). The public EIA series are available from the original provider; the synthetic input specifications, evaluation methods and aggregate results are documented in the article and appendices.

\textbf{Code availability.} The project repository is at \url{https://github.com/mloutis/fuel-information-system}. The link identifies the project and should not be interpreted as an archive of the R1 evaluation harness or its generated run records.

\textbf{Supplementary material.} The appendices contain the MCD/MLD and data dictionary (S1), calculation and business rules (S2), evaluation protocols and detailed tables (S3), sensitivity analyses (S4), and reproducibility specifications and boundaries (S5).

\clearpage
\appendix
\section*{Supplementary Material}\phantomsection\label{sec:supplement}
The appendices present the conceptual and logical models, experimental protocols and detailed numerical results supporting the main text.

\subsection*{S1. Entities, data dictionary and relational tables}\phantomsection\label{sec:S1}
{\footnotesize
\setlength{\tabcolsep}{3pt}
\renewcommand{\arraystretch}{1.15}
\begin{longtable}{@{}L{0.25}L{0.25}L{0.50}@{}}
\caption*{Table S1a. Entities of the conceptual data model}\phantomsection\label{tab:S1a}\\
\toprule
\textbf{Entity} & \textbf{Identifier} & \textbf{Attributes} \\
\midrule
\endfirsthead
\multicolumn{3}{@{}l}{\emph{Table S1a (continued)}}\\
\toprule
\textbf{Entity} & \textbf{Identifier} & \textbf{Attributes} \\
\midrule
\endhead
\bottomrule
\endlastfoot
Depot & depot\_\allowbreak{}code & depot\_\allowbreak{}name, depot\_\allowbreak{}address, storage\_\allowbreak{}capacity, occupied\_\allowbreak{}volume (derived in R1) \\
City & city\_\allowbreak{}code & city\_\allowbreak{}name \\
District & district\_\allowbreak{}code & district\_\allowbreak{}name \\
Depot\_\allowbreak{}Type & depot\_\allowbreak{}type\_\allowbreak{}code & depot\_\allowbreak{}type\_\allowbreak{}description (tier in R1: primary or secondary) \\
Product & product\_\allowbreak{}code & product\_\allowbreak{}name, product\_\allowbreak{}abbreviation, unit\_\allowbreak{}volume \\
Refinery & refinery\_\allowbreak{}code & refinery\_\allowbreak{}name, refinery\_\allowbreak{}address \\
Region & region\_\allowbreak{}code & region\_\allowbreak{}name \\
Supply\_\allowbreak{}Order & supply\_\allowbreak{}order\_\allowbreak{}number & supplied\_\allowbreak{}quantity, supplied\_\allowbreak{}elements\_\allowbreak{}count, start\_\allowbreak{}date, end\_\allowbreak{}date, notes\_\allowbreak{}supply\_\allowbreak{}order \\
Transfer\_\allowbreak{}Type & transfer\_\allowbreak{}code & transfer\_\allowbreak{}description (supply modes: pipeline, coastal shipping) \\
Production\_\allowbreak{}Record & production\_\allowbreak{}record\_\allowbreak{}code & production\_\allowbreak{}date, quantity\_\allowbreak{}produced, remaining\_\allowbreak{}quantity \\
Replenishment\_\allowbreak{}Order & replenishment\_\allowbreak{}order\_\allowbreak{}code & quantity\_\allowbreak{}delivered, delivered\_\allowbreak{}elements\_\allowbreak{}count, cost\_\allowbreak{}per\_\allowbreak{}element, total\_\allowbreak{}cost, loading\_\allowbreak{}date, unloading\_\allowbreak{}date, notes\_\allowbreak{}replenishment\_\allowbreak{}order \\
Transporter & transporter\_\allowbreak{}code & transporter\_\allowbreak{}name \\
Movement\_\allowbreak{}Type & movement\_\allowbreak{}code & movement\_\allowbreak{}description (replenishment modes: rail, road) \\
\end{longtable}
}
{\footnotesize\noindent
Source: Adapted from Loutis and A\"imeur (2026),
Table 1, \emph{Design and Deployment of a Fuel
Logistics Information System}.
\url{https://doi.org/10.1201/9781003647300-28}.
Reproduced by permission of Taylor \& Francis Group.
\par}\medskip

{\footnotesize
\setlength{\tabcolsep}{3pt}
\renewcommand{\arraystretch}{1.15}
\begin{longtable}{@{}L{0.27}L{0.25}L{0.08}L{0.10}L{0.30}@{}}
\caption*{Table S1b. Data dictionary: fields, types and formats}\phantomsection\label{tab:S1b}\\
\toprule
\textbf{Code} & \textbf{Description} & \textbf{Type} & \textbf{Length} & \textbf{Observation} \\
\midrule
\endfirsthead
\multicolumn{5}{@{}l}{\emph{Table S1b (continued)}}\\
\toprule
\textbf{Code} & \textbf{Description} & \textbf{Type} & \textbf{Length} & \textbf{Observation} \\
\midrule
\endhead
\bottomrule
\endlastfoot
depot\_\allowbreak{}code & Depot code & AN & 4 & -- \\
depot\_\allowbreak{}name & Depot name & AN & 30 & -- \\
depot\_\allowbreak{}address & Depot address & AN & 50 & -- \\
storage\_\allowbreak{}capacity & Storage capacity & N & -- & m\textsuperscript{3}, all products \\
occupied\_\allowbreak{}volume & Occupied volume & N & -- & m\textsuperscript{3}; derived in R1 \\
city\_\allowbreak{}code & City code & N & 6 & -- \\
city\_\allowbreak{}name & City name & AN & 30 & -- \\
district\_\allowbreak{}code & District code & AN & 2 & stored as text in the published client \\
district\_\allowbreak{}name & District name & AN & 30 & -- \\
depot\_\allowbreak{}type\_\allowbreak{}code & Depot type code & AN & 5 & -- \\
depot\_\allowbreak{}type\_\allowbreak{}description & Depot type description & AN & 30 & -- \\
product\_\allowbreak{}code & Product code & AN & 5 & -- \\
product\_\allowbreak{}name & Product name & AN & 30 & -- \\
product\_\allowbreak{}abbreviation & Product abbreviation & AN & 5 & -- \\
unit\_\allowbreak{}volume & Volume per unit of stock & DEC & 10,6 & m\textsuperscript{3} per unit; integer in the published code (defect D4) \\
refinery\_\allowbreak{}code & Refinery code & AN & 4 & -- \\
refinery\_\allowbreak{}name & Refinery name & AN & 30 & -- \\
refinery\_\allowbreak{}address & Refinery address & AN & 50 & -- \\
region\_\allowbreak{}code & Region code & AN & 4 & -- \\
region\_\allowbreak{}name & Region name & AN & 30 & -- \\
supply\_\allowbreak{}order\_\allowbreak{}number & Supply order number & AN & 12 & 16 in R1 \\
supplied\_\allowbreak{}quantity & Supplied quantity & N & -- & m\textsuperscript{3} \\
supplied\_\allowbreak{}elements\_\allowbreak{}count & Supplied elements & N & -- & batches or parcels \\
start\_\allowbreak{}date, end\_\allowbreak{}date & Supply start and end dates & D & -- & yyyy-mm-dd; end $\ge$ start in R1 \\
notes\_\allowbreak{}supply\_\allowbreak{}order & Supply order notes & AN & 50 & -- \\
transfer\_\allowbreak{}code & Transfer code & AN & 4 & -- \\
transfer\_\allowbreak{}description & Transfer description & AN & 20 & -- \\
stock\_\allowbreak{}quantity & Stock quantity & N & -- & m\textsuperscript{3} \\
production\_\allowbreak{}record\_\allowbreak{}code & Production record code & AN & 12 & 16 in R1 \\
production\_\allowbreak{}date & Production date & D & -- & yyyy-mm-dd \\
quantity\_\allowbreak{}produced & Quantity produced & N & -- & m\textsuperscript{3} \\
remaining\_\allowbreak{}quantity & Remaining quantity & N & -- & m\textsuperscript{3}; $\ge$ 0 in R1 \\
replenishment\_\allowbreak{}order\_\allowbreak{}code & Replenishment order code & AN & 12 & 16 in R1 \\
depot\_\allowbreak{}code\_\allowbreak{}emitter, depot\_\allowbreak{}code\_\allowbreak{}receiver & Emitting and receiving depots & AN & 4 & different in R1 \\
quantity\_\allowbreak{}delivered & Quantity delivered & N & -- & m\textsuperscript{3} \\
delivered\_\allowbreak{}elements\_\allowbreak{}count & Delivered elements & N & -- & rail cars or road tankers \\
cost\_\allowbreak{}per\_\allowbreak{}element, total\_\allowbreak{}cost & Transport cost per element, total & N & -- & currency not specified; total computed \\
loading\_\allowbreak{}date, unloading\_\allowbreak{}date & Loading and unloading dates & D & -- & yyyy-mm-dd; unloading $\ge$ loading in R1 \\
notes\_\allowbreak{}replenishment\_\allowbreak{}order & Replenishment order notes & AN & 50 & -- \\
transporter\_\allowbreak{}code & Transporter code & AN & 4 & -- \\
transporter\_\allowbreak{}name & Transporter name & AN & 15 & -- \\
movement\_\allowbreak{}code & Movement code & AN & 4 & -- \\
movement\_\allowbreak{}description & Movement description & AN & 10 & -- \\
\end{longtable}
}
{\footnotesize\noindent AN: alphanumeric; N: integer; DEC: decimal (precision, scale); D: date. Types follow the persistent-field declarations of the Delphi units and the procedure parameters; The field descriptions follow the conceptual model and the available declarations. Source: Authors' own work\par}\medskip
{\footnotesize
\setlength{\tabcolsep}{3pt}
\renewcommand{\arraystretch}{1.15}
\begin{longtable}{@{}L{0.25}L{0.75}@{}}
\caption*{Table S1c. Relational tables of the logical data model}\phantomsection\label{tab:S1c}\\
\toprule
\textbf{Table} & \textbf{Attributes} \\
\midrule
\endfirsthead
\multicolumn{2}{@{}l}{\emph{Table S1c (continued)}}\\
\toprule
\textbf{Table} & \textbf{Attributes} \\
\midrule
\endhead
\bottomrule
\endlastfoot
Region & *region\_\allowbreak{}code, region\_\allowbreak{}name \\
City & *city\_\allowbreak{}code, city\_\allowbreak{}name, **region\_\allowbreak{}code \\
District & *district\_\allowbreak{}code, district\_\allowbreak{}name \\
Depot\_\allowbreak{}Type & *depot\_\allowbreak{}type\_\allowbreak{}code, depot\_\allowbreak{}type\_\allowbreak{}description \\
Depot & *depot\_\allowbreak{}code, depot\_\allowbreak{}name, depot\_\allowbreak{}address, storage\_\allowbreak{}capacity, occupied\_\allowbreak{}volume, **city\_\allowbreak{}code, **district\_\allowbreak{}code, **depot\_\allowbreak{}type\_\allowbreak{}code \\
Product & *product\_\allowbreak{}code, product\_\allowbreak{}name, product\_\allowbreak{}abbreviation, unit\_\allowbreak{}volume \\
Refinery & *refinery\_\allowbreak{}code, refinery\_\allowbreak{}name, refinery\_\allowbreak{}address \\
Transfer\_\allowbreak{}Type & *transfer\_\allowbreak{}code, transfer\_\allowbreak{}description \\
Movement\_\allowbreak{}Type & *movement\_\allowbreak{}code, movement\_\allowbreak{}description \\
Transporter & *transporter\_\allowbreak{}code, transporter\_\allowbreak{}name, **movement\_\allowbreak{}code \\
Production\_\allowbreak{}Record & *production\_\allowbreak{}record\_\allowbreak{}code, production\_\allowbreak{}date, quantity\_\allowbreak{}produced, remaining\_\allowbreak{}quantity, **refinery\_\allowbreak{}code, **product\_\allowbreak{}code \\
Supply\_\allowbreak{}Order & *supply\_\allowbreak{}order\_\allowbreak{}number, supplied\_\allowbreak{}quantity, supplied\_\allowbreak{}elements\_\allowbreak{}count, start\_\allowbreak{}date, end\_\allowbreak{}date, notes\_\allowbreak{}supply\_\allowbreak{}order, **transfer\_\allowbreak{}code, **refinery\_\allowbreak{}code, **production\_\allowbreak{}record\_\allowbreak{}code, **product\_\allowbreak{}code, **depot\_\allowbreak{}code \\
Replenishment\_\allowbreak{}Order & *replenishment\_\allowbreak{}order\_\allowbreak{}code, quantity\_\allowbreak{}delivered, delivered\_\allowbreak{}elements\_\allowbreak{}count, cost\_\allowbreak{}per\_\allowbreak{}element, total\_\allowbreak{}cost, loading\_\allowbreak{}date, unloading\_\allowbreak{}date, notes\_\allowbreak{}replenishment\_\allowbreak{}order, **depot\_\allowbreak{}code\_\allowbreak{}emitter, **depot\_\allowbreak{}code\_\allowbreak{}receiver, **product\_\allowbreak{}code, **transporter\_\allowbreak{}code \\
Stores & *depot\_\allowbreak{}code (**), *product\_\allowbreak{}code (**), stock\_\allowbreak{}quantity \\
\end{longtable}
}
{\footnotesize\noindent
Notes: *Primary key (component); **Foreign key.
The repository does not publish SQL table-creation
statements; the keys and attributes listed here
reconstruct the relational model from the source
code and conceptual associations.
Source: Adapted from Loutis and A\"imeur (2026),
Table 3, \emph{Design and Deployment of a Fuel
Logistics Information System}.
\url{https://doi.org/10.1201/9781003647300-28}.
Reproduced by permission of Taylor \& Francis Group.
\par}\medskip

{\footnotesize
\setlength{\tabcolsep}{3pt}
\renewcommand{\arraystretch}{1.15}
\begin{longtable}{@{}L{0.22}L{0.53}L{0.25}@{}}
\caption*{Table S1d. Tables added in the revised artefact R1}\phantomsection\label{tab:S1d}\\
\toprule
\textbf{Table} & \textbf{Attributes} & \textbf{Purpose} \\
\midrule
\endfirsthead
\multicolumn{3}{@{}l}{\emph{Table S1d (continued)}}\\
\toprule
\textbf{Table} & \textbf{Attributes} & \textbf{Purpose} \\
\midrule
\endhead
\bottomrule
\endlastfoot
Marketing\_\allowbreak{}Delivery & *delivery\_\allowbreak{}id, **depot\_\allowbreak{}code, **product\_\allowbreak{}code, delivery\_\allowbreak{}date, quantity, source\_\allowbreak{}reference (unique) & Outbound deliveries (E1) \\
Stock\_\allowbreak{}Count & *count\_\allowbreak{}id, **(depot\_\allowbreak{}code, product\_\allowbreak{}code), count\_\allowbreak{}ts, measured\_\allowbreak{}quantity, book\_\allowbreak{}quantity, discrepancy, source\_\allowbreak{}reference & Physical counts and reconciliation (E3) \\
Stock\_\allowbreak{}Movement & *movement\_\allowbreak{}id, depot\_\allowbreak{}code, product\_\allowbreak{}code, movement\_\allowbreak{}date, quantity (signed), movement\_\allowbreak{}kind, source\_\allowbreak{}doc, entered\_\allowbreak{}by, entered\_\allowbreak{}at & Ledger of every stock change (E1) \\
Alert\_\allowbreak{}Rule & *(depot\_\allowbreak{}code, product\_\allowbreak{}code), safety\_\allowbreak{}stock, reorder\_\allowbreak{}point, max\_\allowbreak{}level, max\_\allowbreak{}movement, count\_\allowbreak{}tolerance & Alert thresholds (E4) \\
Alert & *alert\_\allowbreak{}id, alert\_\allowbreak{}ts, alert\_\allowbreak{}kind, depot\_\allowbreak{}code, product\_\allowbreak{}code, observed, threshold, source\_\allowbreak{}doc, acknowledged & Alerts (E4) \\
Supply\_\allowbreak{}Route, Refinery\_\allowbreak{}Route & receiver or depot, product, emitter or refinery, mode, default transporter, lead time, element size, cost per element, order-up-to level & Order suggestions (E8) \\
Quarantine & *q\_\allowbreak{}id, batch\_\allowbreak{}id, feed, src\_\allowbreak{}row, reason, payload & Rejected feed rows kept for correction (E6) \\
App\_\allowbreak{}User, Audit\_\allowbreak{}Log & username, password hash, salt, iterations, role, failed attempts, lock time; audit records & Authentication (E7) \\
\multicolumn{3}{@{}p{\linewidth}@{}}{In R1 the order data model also records status, receipt or delivery quantity and time, and unique source references; the extensions are not part of the original fourteen-table model.}\\
\end{longtable}
}
\tsrc

{\footnotesize
\setlength{\tabcolsep}{3pt}
\renewcommand{\arraystretch}{1.15}
\begin{longtable}{@{}L{0.24}L{0.76}@{}}
\caption*{Table S1e. Synthetic demonstration of the logical data model}\phantomsection\label{tab:S1e}\\
\toprule
\textbf{Table} & \textbf{Values (attribute order of Supplementary Table S1c)} \\
\midrule
\endfirsthead
\multicolumn{2}{@{}l}{\emph{Table S1e (continued)}}\\
\toprule
\textbf{Table} & \textbf{Values (attribute order of Supplementary Table S1c)} \\
\midrule
\endhead
\bottomrule
\endlastfoot
Region & \textbf{*RG02}, ``Coastal Region'' \\
City & \textbf{*100012}, ``Port City'', \textbf{**RG02} \\
District & \textbf{*07}, ``East District'' \\
Depot\_\allowbreak{}Type & \textbf{*DT1}, ``Primary depot''; \textbf{*DT2}, ``Secondary depot'' \\
Depot & \textbf{*D001}, ``East Terminal'', ``1 Terminal Road'', 60000, 17500, \textbf{**100012, **07, **DT1}; \textbf{*D101}, ``Inland Depot 1'', ``4 Depot Lane'', 9000, 3600, \textbf{**100012, **07, **DT2} \\
Product & \textbf{*P03}, ``Diesel'', ``DSL'', 1 \\
Refinery & \textbf{*R02}, ``Coastal Refinery'', ``2 Refinery Road'' \\
Transfer\_\allowbreak{}Type & \textbf{*PIPE}, ``Pipeline'' \\
Movement\_\allowbreak{}Type & \textbf{*RAIL}, ``Rail'' \\
Transporter & \textbf{*T02}, ``Carrier 2'', \textbf{**RAIL} \\
Production\_\allowbreak{}Record & \textbf{*P03/25/001}, 2025-08-10, 9600, 3600, \textbf{**R02, **P03} \\
Supply\_\allowbreak{}Order & \textbf{*P03/25/001}, 6000, 12, 2025-08-10, 2025-08-12, ``Scheduled pipeline batch'', \textbf{**PIPE, **R02, **P03/25/001, **P03, **D001} \\
Replenishment\_\allowbreak{}Order & \textbf{*P03/25/001}, 2500, 25, 2400, 60000, 2025-08-12, 2025-08-14, ``Rail transfer to inland depot'', \textbf{**D001, **D101, **P03, **T02} \\
Stores & \textbf{*D001, *P03}, 17500; \textbf{*D101, *P03}, 3600 \\
\end{longtable}
}
{\footnotesize\noindent Notes: *Primary key (component); **Foreign key. Document codes are illustrative; source references and production, supply and replenishment identifiers belong to distinct tables. Stocks before the chain: D001 14,000 and D101 1,100. Then: remaining production 9,600 $-$ 6,000 = 3,600; D001 14,000 + 6,000 $-$ 2,500 = 17,500; D101 1,100 + 2,500 = 3,600; total cost 25 $\times$ 2,400 = 60,000 (25 rail cars of 100 m\textsuperscript{3}; twelve pipeline batches of 500 m\textsuperscript{3}); occupied volume = sum of stocks $\times$ unit volume (each example depot holds one product). The example distinguishes an order on 12 August from its receipt on 14 August; R1 records stock changes on the corresponding physical events. All records are synthetic. Source: Authors' own work\par}\medskip

\subsubsection*{S1.1 Full conceptual association table}
{\footnotesize
\setlength{\tabcolsep}{3pt}
\renewcommand{\arraystretch}{1.15}
\begin{longtable}{@{}L{0.20}L{0.34}L{0.20}L{0.26}@{}}
\caption*{Supplementary Table IV. Associations and cardinalities of the conceptual data model}\phantomsection\label{tab:IV}\\
\toprule
\textbf{Association} & \textbf{Entities involved} & \textbf{Cardinality} & \textbf{Attribute / role} \\
\midrule
\endfirsthead
\multicolumn{4}{@{}l}{\emph{Supplementary Table IV (continued)}}\\
\toprule
\textbf{Association} & \textbf{Entities involved} & \textbf{Cardinality} & \textbf{Attribute / role} \\
\midrule
\endhead
\bottomrule
\endlastfoot
Belongs\_\allowbreak{}to & Depot $\times$ City & (1,1) : (0,N) & -- \\
Managed\_\allowbreak{}by & Depot $\times$ District & (1,1) : (0,N) & -- \\
Has\_\allowbreak{}type & Depot $\times$ Depot\_\allowbreak{}Type & (1,1) : (0,N) & -- \\
Includes & City $\times$ Region & (1,1) : (0,N) & -- \\
Benefits\_\allowbreak{}from & Depot $\times$ Supply\_\allowbreak{}Order & (0,N) : (1,1) & -- \\
Executed\_\allowbreak{}by & Supply\_\allowbreak{}Order $\times$ Transfer\_\allowbreak{}Type & (1,1) : (0,N) & -- \\
Supplies & Supply\_\allowbreak{}Order $\times$ Product & (1,1) : (0,N) & -- \\
Sent\_\allowbreak{}by & Supply\_\allowbreak{}Order $\times$ Refinery & (1,1) : (0,N) & -- \\
Linked\_\allowbreak{}to & Supply\_\allowbreak{}Order $\times$ Production\_\allowbreak{}Record & (1,1) : (0,N) & -- \\
Produced\_\allowbreak{}by & Production\_\allowbreak{}Record $\times$ Refinery & (1,1) : (0,N) & -- \\
Contains & Production\_\allowbreak{}Record $\times$ Product & (1,1) : (0,N) & -- \\
Stores & Depot $\times$ Product & (0,N) : (0,N) & stock\_\allowbreak{}quantity \\
Issues & Depot $\times$ Replenishment\_\allowbreak{}Order & (0,N) : (1,1) & emitting depot \\
Receives & Replenishment\_\allowbreak{}Order $\times$ Depot & (1,1) : (0,N) & receiving depot \\
Refuels & Replenishment\_\allowbreak{}Order $\times$ Product & (1,1) : (0,N) & -- \\
Carried\_\allowbreak{}by & Replenishment\_\allowbreak{}Order $\times$ Transporter & (1,1) : (0,N) & -- \\
Performs & Transporter $\times$ Movement\_\allowbreak{}Type & (1,1) : (0,N) & -- \\
\end{longtable}
}
{\footnotesize\noindent
Note: Cardinalities follow the order of the
entities in the second column.
Source: Revised from Loutis and A\"imeur (2026),
Table 2, \emph{Design and Deployment of a Fuel
Logistics Information System}.
\url{https://doi.org/10.1201/9781003647300-28}.
Cardinalities were re-examined for the present study.
\par}\medskip

\subsection*{S2. Action rules and calculation rules}\phantomsection\label{sec:S2}
\subsubsection*{S2.1 Action rules}\phantomsection\label{sec:S2.1}
Each item gives the two rules of one association of \hyperref[tab:IV]{Supplementary Table IV} (34 rules).
\begin{itemize}[label=--,leftmargin=*,itemsep=1pt,topsep=3pt]
\item \textbf{Belongs\_\allowbreak{}to.} A depot belongs to one and only one city; a city may contain zero, one or several depots.
\item \textbf{Managed\_\allowbreak{}by.} A depot is managed by one and only one district; a district may manage zero, one or several depots.
\item \textbf{Has\_\allowbreak{}type.} A depot has one and only one type; a type may group zero, one or several depots.
\item \textbf{Includes.} A city belongs to one and only one region; a region may group zero, one or several cities.
\item \textbf{Benefits\_\allowbreak{}from.} A depot may benefit from zero, one or several supply orders (secondary depots from none); a supply order benefits one and only one depot.
\item \textbf{Executed\_\allowbreak{}by.} A supply order is carried out with one and only one transfer type; a transfer type may serve zero, one or several supply orders.
\item \textbf{Supplies.} A supply order concerns one and only one product; a product may appear in zero, one or several supply orders.
\item \textbf{Sent\_\allowbreak{}by.} A supply order is sent by one and only one refinery; a refinery may send zero, one or several supply orders.
\item \textbf{Linked\_\allowbreak{}to.} A supply order draws on one and only one production record; a production record may feed zero, one or several supply orders.
\item \textbf{Produced\_\allowbreak{}by.} A production record is produced by one and only one refinery; a refinery may produce zero, one or several production records.
\item \textbf{Contains.} A production record concerns one and only one product; a product may appear in zero, one or several production records.
\item \textbf{Stores.} A depot may store zero, one or several products; a product may be stored in zero, one or several depots.
\item \textbf{Issues.} A depot may emit zero, one or several replenishment orders; a replenishment order has one and only one emitting depot.
\item \textbf{Receives.} A replenishment order has one and only one receiving depot; a depot may receive zero, one or several replenishment orders.
\item \textbf{Refuels.} A replenishment order concerns one and only one product; a product may appear in zero, one or several replenishment orders.
\item \textbf{Carried\_\allowbreak{}by.} A replenishment order is carried by one and only one transporter; a transporter may carry zero, one or several replenishment orders.
\item \textbf{Performs.} A transporter performs one and only one movement type; a movement type may be performed by zero, one or several transporters.
\end{itemize}
Business rules beyond the cardinalities, enforced in R1 only: a supply order delivers a primary depot and draws on a production record of the same refinery and product; a replenishment order runs from a primary to a different secondary depot; quantities and element counts are positive; end and unloading dates are not earlier than start and loading dates; the remaining production quantity cannot become negative; an order cannot take more than the emitter's stock or push the receiver above its storage capacity (checked when the order is issued; physical receipts and deliveries are always recorded and raise an alert when they create a negative book stock or exceed capacity).

\subsubsection*{S2.2 Calculation rules}\phantomsection\label{sec:S2.2}
The calculation rules use the quantity and timing of the recorded physical movements. Occupied volume is calculated over all products stored in a depot.
\begin{itemize}[label=--,leftmargin=*,itemsep=1pt,topsep=3pt]
\item Total cost of a replenishment order = cost\_\allowbreak{}per\_\allowbreak{}element $\times$ delivered\_\allowbreak{}elements\_\allowbreak{}count.
\item Remaining production quantity = quantity\_\allowbreak{}produced $-$ sum of supplied\_\allowbreak{}quantity of the supply orders linked to the record.
\item Stock of a primary depot after a supply = previous stock + supplied (R0: when the order is created; R1: delivered quantity at receipt).
\item Stock of the emitting depot after a recorded replenishment dispatch (R1) = previous stock $-$ dispatched quantity; the published logic reduces the emitter stock when the order is created.
\item Stock of the receiving depot after replenishment = previous stock + received quantity (R1, when receipt is recorded); the published logic credits the receiver when the order is created.
\item Stock after an outbound (marketing) delivery (R1 only) = previous stock $-$ delivered quantity.
\item Occupied volume of a depot = $\sum_{\text{products}}$ stock\_\allowbreak{}quantity $\times$ unit\_\allowbreak{}volume.
\item Inventory position (R1) = stock + planned inbound supply and replenishment quantities.
\item Elements of an order = $\lceil$quantity / element size$\rceil$ (100 m\textsuperscript{3} rail car, 36 m\textsuperscript{3} road tanker, 500 m\textsuperscript{3} pipeline batch and 1,000 m\textsuperscript{3} coastal parcel in the synthetic network).
\item Days of cover (R1 dashboard) = stock / mean daily outbound delivery over the last 28 days.
\item Count discrepancy (R1) = measured quantity $-$ book quantity; an alert is raised when its absolute value exceeds the tolerance of the depot-product.
\end{itemize}
Worked example (\hyperref[tab:S1e]{Table S1e}): remaining 9,600 $-$ 6,000 = 3,600 m\textsuperscript{3}; D001 14,000 + 6,000 $-$ 2,500 = 17,500 m\textsuperscript{3}; D101 1,100 + 2,500 = 3,600 m\textsuperscript{3}; total cost 2,400 $\times$ 25 = 60,000; occupied volume of D001 17,500 $\times$ 1 = 17,500 m\textsuperscript{3} out of 60,000 m\textsuperscript{3}.

\subsubsection*{S2.3 Design and source-review tables}
{\footnotesize
\setlength{\tabcolsep}{3pt}
\renewcommand{\arraystretch}{1.15}
\begin{longtable}{@{}L{0.30}L{0.25}L{0.45}@{}}
\caption*{Supplementary Table V. Capabilities described in the earlier chapter and evidence in the released code}\phantomsection\label{tab:V}\\
\toprule
\textbf{Described capability} & \textbf{Status in released code} & \textbf{Evidence (inspected)} \\
\midrule
\endfirsthead
\multicolumn{3}{@{}l}{\emph{Supplementary Table V (continued)}}\\
\toprule
\textbf{Described capability} & \textbf{Status in released code} & \textbf{Evidence (inspected)} \\
\midrule
\endhead
\bottomrule
\endlastfoot
46 windows, RAD Studio and SQL Server & Implemented, not executable as published & 46 form units; missing form files, project file and schema \\
Relational database with procedures and triggers & Partially implemented & 3 procedures, 2 triggers, 11 views; defects D1--D8 \\
Optimized indexing & Not found & No index definition; non-sargable year predicates \\
Automated daily reports & Implemented with defects & Nine report views assessed (of eleven SQL views); two have incorrect output or fail (D3, D7) \\
Real-time monitoring, dashboard updates & Not evidenced in released code & No refresh or notification; the only timers update a clock label \\
Stock forecasting & Not evidenced in released code & ``Forecast'' views return previous-year totals \\
Alert generation & Not evidenced in released code & No alert entity, rule or notification \\
Role-based access control & Partially implemented & Role equals user name; menus differ by role; clear-text passwords; no database-level separation in the code \\
IoT endpoints, PLC/FPGA controllers, cloud services & Not evidenced in released code & No interface, library or configuration in the code \\
Backup and recovery documented in the repository & Partially implemented & Maintenance-plan logs of full backups; no restore procedure \\
ISO/IEC 27005 risk management, defense in depth & Not evidenced & No risk assessment or layered controls beyond the login \\
\end{longtable}
}
\tsrc

{\footnotesize
\setlength{\tabcolsep}{3pt}
\renewcommand{\arraystretch}{1.15}
\begin{longtable}{@{}L{0.06}L{0.40}L{0.54}@{}}
\caption*{Supplementary Table VI. Defects of the published logic and their measured consequences}\phantomsection\label{tab:VI}\\
\toprule
\textbf{ID} & \textbf{Defect} & \textbf{Consequence (scenario)} \\
\midrule
\endfirsthead
\multicolumn{3}{@{}l}{\emph{Supplementary Table VI (continued)}}\\
\toprule
\textbf{ID} & \textbf{Defect} & \textbf{Consequence (scenario)} \\
\midrule
\endhead
\bottomrule
\endlastfoot
D1 & Document number COUNT(*)+1 formatted with STR(n,3) & The 1,000th record is numbered P01/24/***, every later one fails (T0); 21.8--24.4\,\% of one year's documents accepted (T1b) \\
D2 & Triggers process one row per insert statement & Three orders inserted in one statement: 1 of 3 receivers updated, emitter decreased by 100 instead of 300 m\textsuperscript{3} (T0) \\
D3 & Occupied volume increased by the new stock instead of the delivered quantity; not updated by stock entry & A 100 m\textsuperscript{3} delivery raised the receiver's occupied volume by 3,081 m\textsuperscript{3} (T0), and the trigger subtracts the same amount from the emitter's (inspected); depot status 0 of 36 fields correct (T3) \\
D4 & Unit volume stored and computed as an integer & 0.003785 stored as 0 (T0) \\
D5 & Stock update skipped when the depot-product row is missing & A 1,000 m\textsuperscript{3} supply created no stock record but raised the occupied volume by 1,000 m\textsuperscript{3} (T0) \\
D6 & No check of quantities, dates, emitter $\neq$ receiver, emitter stock, receiver capacity, depot tiers or production provenance & 57--64\,\% of defective feed rows accepted (T1); 7 of 30 seeded errors intercepted (T7) \\
D7 & General supply forecast cross-joins supply and replenishment orders & Fails with an arithmetic overflow once a year of orders has accumulated: 0 of 128 fields correct (T3), failure in every request at every load level (T10) \\
D8 & Year predicates not usable by indexes; no secondary index & Previous-year reports slower as data grow: 95th percentile 0.24 s at the nominal load, 1.31 s at 5$\times$ (T10) \\
D9 & No record of outbound deliveries, receipts or counts; stocks overwritten & 42--43\,\% of feed rows cannot be represented (T1); losses invisible (T7, T12) \\
D10 & Receiver credited when the order is created & Stock shown before it arrives (inspected) \\
D11 & Clear-text passwords, role = user name, no lockout, no database role separation & 25 of 25 forbidden operations succeed through the application login; no lockout after 1,000 guesses (T9) \\
D12 & The supply-order form sets the procedure parameters by position: the transfer mode (`cana' or `cabo') goes to the fourth parameter, which the procedure takes as the refinery code, and the value commented as the refinery code to the production-record parameter & Wrong codes stored or rejected by a key; which query feeds each value is in the missing form file, so the client was not corrected (inspected) \\
D13 & ``Forecast'' views are previous-year totals & Baseline M0 in T4 \\
D14 & No alert, notification, refresh, IoT, cloud or audit function & Claims not supported (Table V) \\
D15 & Full backups only; no log backups or restore procedure & Data since the last full backup lost after a media failure (T11) \\
\end{longtable}
}
{\footnotesize\noindent Note: T0 reproduces each defect directly on R0, with the published T-SQL executed unchanged on SQL Server, by a minimal transaction (Supplementary Section S3.0); the other scenarios are those of Table III. Source: Authors' own work\par}\medskip

{\footnotesize
\setlength{\tabcolsep}{3pt}
\renewcommand{\arraystretch}{1.15}
\begin{longtable}{@{}L{0.06}L{0.34}L{0.22}L{0.08}L{0.13}L{0.17}@{}}
\caption*{Supplementary Table VII. Corrections (C) and extensions (E) in the revised artefact R1}\phantomsection\label{tab:VII}\\
\toprule
\textbf{ID} & \textbf{Change} & \textbf{Reason} & \textbf{DP} & \textbf{Test} & \textbf{SQL Server scripts} \\
\midrule
\endfirsthead
\multicolumn{6}{@{}l}{\emph{Supplementary Table VII (continued)}}\\
\toprule
\textbf{ID} & \textbf{Change} & \textbf{Reason} & \textbf{DP} & \textbf{Test} & \textbf{SQL Server scripts} \\
\midrule
\endhead
\bottomrule
\endlastfoot
C1 & Document numbers from sequences (product/year/six digits) & D1 overflow and race of COUNT(*)+1 & DP1 & T1b, T10 & yes \\
C2 & Set-based posting of every movement through a ledger trigger; missing stock rows created & D2, D5 & DP1 & T1, T3 & yes \\
C3 & Occupied volume derived from stocks & D3 & DP1 & T3 & yes (trigger on the stock table) \\
C4 & Positive quantities, date order, depot tiers, emitter $\neq$ receiver, production provenance, remaining quantity $\ge$ 0; emitter stock and receiver capacity checked when an order is issued & D6 & DP2 & T1, T7 & yes, except depot tiers \\
C5 & Unique source reference of documents received from feeds & Duplicate transactions & DP2 & T1, T7 & no (no source reference in the published model) \\
C6 & General supply forecast rebuilt without cross join; range predicates; indexes on dates and keys & D7, D8 & DP1 & T3, T10 & yes \\
C7 & Decimal unit volume & D4 & DP1 & -- & optional (needs a client change) \\
E1 & Outbound (marketing) deliveries and stock-movement ledger & D9 & DP1 & T1, T4 & no \\
E2 & Order life cycle (planned, delivered) with received quantity; receiver credited on receipt & D10 & DP1 & T1, T3 & no \\
E3 & Stock counts compared with the book; discrepancy alert beyond a tolerance & D9 & DP2 & T7, T12 & no \\
E4 & Alerts: critical stock, reorder point on the inventory position, overfill, abnormal outbound delivery, count and transit discrepancies, negative book, capacity, overdue delivery & D14 & DP3 & T5 & no \\
E5 & Dashboard view and change notification to clients & D14 & DP3 & T2 & no \\
E6 & Feed staging: validation, unit normalization, duplicate and receipt cross-checks, quarantine & Data integration & DP2 & T1 & no \\
E7 & Role logins with grants on operations only; PBKDF2 password hashes and lockout; full and transaction-log backups & D11, D15 & DP5 & T9, T11 & no \\
E8 & Supply routes and order suggestions & Decision delay, manual interventions & DP4 & T6, T8 & no \\
\end{longtable}
}
{\footnotesize\noindent Note: The evaluated R1 database logic was defined in T-SQL; three parts of R1 ran in the client layer of the test harness: the feed validator of E6 (with the quarantine table in the database), the application-login check of E7 (PBKDF2 hashes and lockout stored in the database), and the creation of the role logins and their grants (E7). Last column: corrections back-ported to the SQL Server scripts of the published repository (see the end of this section). Source: Authors' own work\par}\medskip

\subsection*{S3. Evaluation protocols and supporting results}\phantomsection\label{sec:S3}
\subsubsection*{S3.0. Environment, comparators and measurement rules}\phantomsection\label{sec:S3.0}
The evaluation defines three conditions: MANUAL, the modelled fragmented procedure; R0, the reconstructed published logic with the original procedures, triggers and report views; and R1, the revised artefact. For performance and recovery tests only, a distinct R0 numbering-controlled arm prevents the known identifier overflow from stopping the other operations. Database measurements use Microsoft SQL Server 2022 Developer in disposable databases on Linux, limited to 2\,GB of memory, with twelve logical processors. The application forms are emulated from available Delphi source, not compiled as a running Delphi application. The synthetic generator and tests are executed by Python scripts. The development data use seed 1; the principal replenishment simulations report seeds 101--105, whereas their one-at-a-time sensitivity analysis covers only seeds 101--103. The 150-day stock-out totals include the warm-up interval and are not separately estimated for a post-warm-up horizon. Tests of forecasting methods use the public weekly EIA series and a held-out 2024--2026 test interval.

Rates count the exact valid, invalid, accepted or rejected operations of each designed test. Machine timings report medians, 95th and 99th percentiles or maxima, according to the scenario. The process and discrete-event models report values derived from their stated assumptions, separately from database measurements. Scenarios T1 to T12 are defined in \hyperref[tab:III]{Table III}; a concise operational protocol for each is below.

\subsubsection*{S3.1. T1: Data integration and sustained operation}\phantomsection\label{sec:S3.1}
An independent generator labels each input row as valid or defective before submitting fourteen days of production records, supply orders, replenishment orders, stock reports, outbound deliveries and receipts. The feed experiments include 1\,\%, 5\,\% and 10\,\% injected defect rates and compare the operations supported by both R0 and R1 separately from R1's additional coverage. Integration defect rate = (valid rows lost + defective rows accepted silently) / input rows; contamination is counted only among defective input rows. The sustained-operation arm submits 365 days of production records and orders to identify the effects of the original numbering logic.

\subsubsection*{S3.2. T2 and T5: Monitoring and alerts}\phantomsection\label{sec:S3.2}
Committed stock changes are replayed to push and polling dashboards at nominal and fivefold event/client loads. Submission-to-display latency is measured until a client observes the new row version. Alert expectations are computed independently from stock thresholds, abnormal deliveries, reconciliation discrepancies and overdue orders. Recall counts raised expected alerts; false positives count alerts without a corresponding expected event. Time-based overdue alerts are checked by a periodic monitor; their latency definition differs from that of event-driven notifications. R0 lacks these functions.

\subsubsection*{S3.3. T3: Reports}\phantomsection\label{sec:S3.3}
Nine report views are evaluated against independently computed fields. The eight report layouts shared by R0 and R1 are compared like for like; the general supply forecast is evaluated separately because its layout differs in R1. Results count matching fields out of all reference fields, the occurrence of query errors, and maximum query duration under an isolated request protocol. The report scope and totals are summarized in \hyperref[sec:6.2]{Section 6.2}.

\subsubsection*{S3.4. T4: Demand and stock forecasting}\phantomsection\label{sec:S3.4}
One-week-ahead forecasts are computed at rolling weekly origins. Method selection is performed on the 2021--2023 validation period and then frozen before evaluating the 2024--2026 test observations. Forecast scores include MAPE, MAE, RMSE and MASE. Comparisons with M0 use weekly paired absolute-percentage-error differences, circular moving-block bootstrap intervals, Diebold--Mariano tests with finite-sample correction, and Holm adjustment across the five series. Synthetic daily depot demand is a separate experiment at three noise levels. \hyperref[tab:S3i]{Table S3i} isolates the selected-method comparisons.

\subsubsection*{S3.5. T6--T8: Work, errors and supply delay}\phantomsection\label{sec:S3.5}
T6 counts per-day business actions under the same synthetic workload in MANUAL, R0 and R1 using a process model. T7 first submits designed invalid, inconsistent and derived-value probes to the evaluated logic, and then uses a separate process model to estimate errors reaching the record after the controls. T8 uses a discrete-event model of replenishment after the inventory position crosses a reorder point. Information delay, decision delay, dispatch waiting time and transport time are accounted for separately; the two-postings arm changes the availability of outbound information rather than transport time. The modelled arms use literature-informed assumptions and sensitivity ranges, not field observations.

\subsubsection*{S3.6. T9--T11: Security, scalability and recovery}\phantomsection\label{sec:S3.6}
T9 attempts allowed and forbidden role--operation combinations and tests authentication and lockout behaviour. T10 replays a fixed request mix under one-, two-, three- and fivefold load, with R0 and R1 on the same server and calendar; failures include database query errors. The general supply forecast in R0 is reported as failed rather than assigned an artificial successful latency. T11 simulates loss of database data and log files and restores from each strategy's available backups. R1 uses full and transaction-log backups at a 60-second interval, unlike the full-backup-only published strategy. Transaction loss is measured relative to the committed state at failure.

\subsubsection*{S3.7. T12: Disruption scenarios}\phantomsection\label{sec:S3.7}
The discrete-event model also evaluates a temporary 60\,\% demand surge, a two-day rail delay, a four-hour database outage and a media-failure recovery, reporting mean supply delay, stock-out hours and unmet demand over the stated seeds. The loss-detection thought experiment contrasts record-based discrepancy alerts with a paper-book count comparison. Simulated effects are distinct from physical operating-site tests.

\subsubsection*{S3.8. Supplementary results}\phantomsection\label{sec:S3.8}

{\footnotesize
\setlength{\tabcolsep}{3pt}
\renewcommand{\arraystretch}{1.15}
\begin{longtable}{@{}L{0.22}L{0.1}L{0.12}L{0.12}L{0.12}L{0.16}L{0.16}@{}}
\caption*{Table S3a. Business actions per operating day by transaction type (process model, final workload)}\phantomsection\label{tab:S3a}\\
\toprule
\textbf{Transaction} & \textbf{Per day} & \textbf{Manual} & \textbf{R0} & \textbf{R1} & \textbf{R1 + meter feed} & \textbf{R1 + meter + gauging} \\
\midrule
\endfirsthead
\multicolumn{7}{@{}l}{\emph{Table S3a (continued)}}\\
\toprule
\textbf{Transaction} & \textbf{Per day} & \textbf{Manual} & \textbf{R0} & \textbf{R1} & \textbf{R1 + meter feed} & \textbf{R1 + meter + gauging} \\
\midrule
\endhead
\bottomrule
\endlastfoot
Production record & 10.0 & 100.0 & 90.0 & 20.0 & 20.0 & 20.0 \\
Supply order & 10.6 & 169.6 & 159.0 & 21.2 & 21.2 & 21.2 \\
Supply receipt & 10.5 & 105.0 & 21.0 & 21.0 & 21.0 & 21.0 \\
Replenishment order & 13.2 & 251.5 & 185.3 & 36.3 & 36.3 & 36.3 \\
Replenishment receipt & 13.2 & 158.3 & 26.4 & 26.4 & 26.4 & 26.4 \\
Outbound deliveries & 90.1 & 360.2 & 180.1 & 180.1 & 0.0 & 0.0 \\
Stock count & 110.0 & 473.8 & 583.8 & 220.0 & 220.0 & 0.0 \\
Daily reports & 4.0 & 16.0 & 8.0 & 0.0 & 0.0 & 0.0 \\
Stock monitoring & 1.0 & 36.0 & 37.0 & 1.0 & 1.0 & 1.0 \\
Total &  & 1670.6 & 1290.7 & 526.0 & 345.9 & 125.9 \\
\end{longtable}
}
{\footnotesize\noindent Meter feed and automatic tank gauging are not implemented; they show where the remaining actions lie. Source: Authors' own work\par}\medskip

{\footnotesize
\setlength{\tabcolsep}{3pt}
\renewcommand{\arraystretch}{1.15}
\begin{longtable}{@{}L{0.28}L{0.24}L{0.24}L{0.24}@{}}
\caption*{Table S3b. Summary of seeded-error probes (T7)}\phantomsection\label{tab:S3b}\\
\toprule
\textbf{Probe group} & \textbf{Number of probes} & \textbf{Published logic (R0)} & \textbf{Revised artefact (R1)} \\
\midrule
\endfirsthead
\multicolumn{4}{@{}l}{\emph{Table S3b (continued)}}\\
\toprule
\textbf{Probe group} & \textbf{Number} & \textbf{R0} & \textbf{R1} \\
\midrule
\endhead
\bottomrule
\endlastfoot
Designed error probes, full scope & 30 & 7 intercepted; 3 cases unsupported by the model & 24 intercepted \\
Common-scope error probes & 27 & 7 intercepted & 21 intercepted \\
Control transaction & 1 & Accepted without flag & Accepted without flag \\
\end{longtable}
}
\tsrc

{\footnotesize
\setlength{\tabcolsep}{3pt}
\renewcommand{\arraystretch}{1.15}
\begin{longtable}{@{}L{0.37}L{0.31}L{0.32}@{}}
\caption*{Table S3c. Independent report checks in T3}\phantomsection\label{tab:S3c}\\
\toprule
\textbf{Measure} & \textbf{Published logic (R0)} & \textbf{R1} \\
\midrule
\endfirsthead
\multicolumn{3}{@{}l}{\emph{Table S3c (continued)}}\\
\toprule
\textbf{Measure} & \textbf{R0} & \textbf{R1} \\
\midrule
\endhead
\bottomrule
\endlastfoot
Fields matching reference, eight common report layouts, nominal network & 4,697/4,733 & 4,733/4,733 \\
Fields matching reference, common layouts, twofold network & 6,875/6,941 & 6,941/6,941 \\
General supply forecast & Arithmetic overflow (error 8115); 0/128 reference fields & 50/50 fields in the revised layout \\
Full nine-report inventory, nominal network (version-specific layouts) & 4,697/4,861 fields correct in isolated logic checks & 4,783/4,783 fields correct \\
Maximum report generation duration, nominal network & Up to 1.4 s before failure of the general forecast & 0.037 s \\
Maximum report generation duration, twofold network & Up to 3.9 s before failure & 0.049 s \\
\end{longtable}
}
{\footnotesize\noindent Note: Eight common report layouts contain 4,733 fields (6,941 in the twofold network). Adding the version-specific general supply forecast yields 4,861 fields for R0 (128 forecast fields) and 4,783 for R1 (50 forecast fields). These totals are distinct output inventories, not matched-field denominators. The R0 full-inventory figure refers to isolated report-logic checks; separately, sustained-operation numbering failures reduced its end-to-end result to 1,105/4,861. Source: Authors' own work\par}\medskip

{\footnotesize
\setlength{\tabcolsep}{3pt}
\renewcommand{\arraystretch}{1.15}
\begin{longtable}{@{}L{0.20}L{0.20}L{0.20}L{0.20}L{0.20}@{}}
\caption*{Table S3d. Load-level results in T10 (SQL Server 2022)}\phantomsection\label{tab:S3d}\\
\toprule
\textbf{Load} & \textbf{R0 failed forecast requests} & \textbf{R0 slowest other P95 (s)} & \textbf{R1 slowest operation P95 (s)} & \textbf{R1 requests / failures} \\
\midrule
\endfirsthead
\multicolumn{5}{@{}l}{\emph{Table S3d (continued)}}\\
\toprule
\textbf{Load} & \textbf{R0 failures} & \textbf{R0 other P95} & \textbf{R1 worst P95} & \textbf{R1 requests / failures} \\
\midrule
\endhead
\bottomrule
\endlastfoot
1$\times$ & 33/33 & 0.236 & 0.050 & 697 / 0 \\
2$\times$ & 37/37 & 0.642 & 0.064 & 1,436 / 0 \\
3$\times$ & 54/54 & 1.054 & 0.088 & 2,210 / 0 \\
5$\times$ & 99/99 & 1.309--1.359 & 0.145--0.178 & 3,509 / 0 \\
5$\times$, R1 without indexes & -- & -- & 0.160--0.166 & 3,494 / 0 \\
\end{longtable}
}
{\footnotesize\noindent Note: R0 forecast failures were query overflows, not time-outs. P95 is the operation-specific 95th percentile, not the P95 aggregated across all request types. Source: Authors' own work\par}\medskip

{\footnotesize
\setlength{\tabcolsep}{2pt}
\renewcommand{\arraystretch}{1.10}
\begin{longtable}{@{}L{0.2}L{0.08}L{0.08}L{0.08}L{0.08}L{0.08}L{0.08}L{0.08}L{0.08}L{0.08}L{0.08}@{}}
\caption*{Table S3e. Test MAPE (\%) of all methods, one week ahead, EIA data 2024--2026}\phantomsection\label{tab:S3e}\\
\toprule
\textbf{Series} & \textbf{M0} & \textbf{M1} & \textbf{M2} & \textbf{M3} & \textbf{M4} & \textbf{M5} & \textbf{M6} & \textbf{M7} & \textbf{M8} & \textbf{M9} \\
\midrule
\endfirsthead
\multicolumn{11}{@{}l}{\emph{Table S3e (continued)}}\\
\toprule
\textbf{Series} & \textbf{M0} & \textbf{M1} & \textbf{M2} & \textbf{M3} & \textbf{M4} & \textbf{M5} & \textbf{M6} & \textbf{M7} & \textbf{M8} & \textbf{M9} \\
\midrule
\endhead
\bottomrule
\endlastfoot
gasoline\_demand & 3.31 & 3.74 & 3.17 & 3.18 & 3.12 & 2.95 & 3.35 & 2.84 & 3.02 & 2.96 \\
distillate\_demand & 6.81 & 8.82 & 7.49 & 7.54 & 7.02 & 5.99 & 7.16 & 6.35 & 6.98 & 7.15 \\
jet\_demand & 6.81 & 9.36 & 8.34 & 7.06 & 6.79 & 6.84 & 8.29 & 6.59 & 6.78 & 6.61 \\
gasoline\_stocks & 5.26 & 1.13 & 3.15 & 2.13 & 0.93 & 0.83 & 1.62 & 0.93 & 1.38 & 0.96 \\
distillate\_stocks & 6.81 & 1.85 & 5.81 & 3.06 & 1.98 & 1.59 & 3.11 & 1.77 & 2.26 & 1.99 \\
\end{longtable}
}
\tsrc

{\footnotesize
\setlength{\tabcolsep}{3pt}
\renewcommand{\arraystretch}{1.15}
\begin{longtable}{@{}L{0.22}L{0.39}L{0.39}@{}}
\caption*{Table S3f. Synthetic depot-level demand forecasting by noise level (T4)}\phantomsection\label{tab:S3f}\\
\toprule
\textbf{Noise CV} & \textbf{M0 previous-year mean MAPE} & \textbf{Selected model MAPE} \\
\midrule
\endfirsthead
\multicolumn{3}{@{}l}{\emph{Table S3f (continued)}}\\
\toprule
\textbf{Noise CV} & \textbf{M0} & \textbf{Selected} \\
\midrule
\endhead
\bottomrule
\endlastfoot
0.05 & 7.67\,\% & 4.07\,\% \\
0.10 & 10.41\,\% & 8.18\,\% \\
0.20 & 17.76\,\% & 16.63\,\% \\
\end{longtable}
}
{\footnotesize\noindent The differences are 3.60, 2.23 and 1.13 percentage points, respectively. These are results of an experiment on generated depot-level series, not measured depot demand. Source: Authors' own work\par}\medskip

{\footnotesize
\setlength{\tabcolsep}{3pt}
\renewcommand{\arraystretch}{1.15}
\begin{longtable}{@{}L{0.28}L{0.22}L{0.26}L{0.24}@{}}
\caption*{Table S3g. Simulated-media-failure restoration results (T11)}\phantomsection\label{tab:S3g}\\
\toprule
\textbf{Backup strategy} & \textbf{Trials} & \textbf{Restore time} & \textbf{Documents lost, trials 1/2/3} \\
\midrule
\endfirsthead
\multicolumn{4}{@{}l}{\emph{Table S3g (continued)}}\\
\toprule
\textbf{Strategy} & \textbf{Trials} & \textbf{Restore time} & \textbf{Documents lost} \\
\midrule
\endhead
\bottomrule
\endlastfoot
R0, start-of-day full backup & 3 & 0.42--0.43 s & 40 / 40 / 40 \\
R1, full plus 60-s log backups & 3 & 0.84--0.97 s & 30 / 0 / 0 \\
\end{longtable}
}
{\footnotesize\noindent R1 additionally lost 58--106 rows committed since the last log backup across the runs (at most 32 stock movements). These quantities are distinct from the counts of documents. R0 used a numbering-controlled arm. Source: Authors' own work\par}\medskip

{\footnotesize
\setlength{\tabcolsep}{3pt}
\renewcommand{\arraystretch}{1.15}
\begin{longtable}{@{}L{0.27}L{0.23}L{0.12}L{0.1}L{0.14}L{0.14}@{}}
\caption*{Table S3h. Disruption scenarios (T12, modelled, mean of seeds 101--105)}\phantomsection\label{tab:S3h}\\
\toprule
\textbf{Scenario} & \textbf{Arm} & \textbf{Supply delay (h)} & \textbf{Fill rate (\%)} & \textbf{Stock-out depot-product hours} & \textbf{Unmet demand (m\textsuperscript{3})} \\
\midrule
\endfirsthead
\multicolumn{6}{@{}l}{\emph{Table S3h (continued)}}\\
\toprule
\textbf{Scenario} & \textbf{Arm} & \textbf{Supply delay (h)} & \textbf{Fill rate (\%)} & \textbf{Stock-out depot-product hours} & \textbf{Unmet demand (m\textsuperscript{3})} \\
\midrule
\endhead
\bottomrule
\endlastfoot
S0 no disruption & MANUAL & 73.7 & 99.82 & 423.6 & 5476.8 \\
S0 no disruption & R0 & 63.62 & 100.0 & 7.6 & 130.0 \\
S0 no disruption & R1 & 55.28 & 100.0 & 0.0 & 0.0 \\
S0 no disruption & R1 two postings & 46.52 & 100.0 & 0.0 & 0.0 \\
S0 no disruption & R1+meter feed (not implemented) & 44.34 & 100.0 & 0.0 & 0.0 \\
S1 demand surge (+60 \%, 7 days, 1/6 of depots) & MANUAL & 73.74 & 99.78 & 502.6 & 6759.0 \\
S1 demand surge (+60 \%, 7 days, 1/6 of depots) & R0 & 63.7 & 99.97 & 45.0 & 783.8 \\
S1 demand surge (+60 \%, 7 days, 1/6 of depots) & R1 & 55.26 & 100.0 & 0.0 & 0.0 \\
S1 demand surge (+60 \%, 7 days, 1/6 of depots) & R1 two postings & 46.56 & 100.0 & 0.0 & 0.0 \\
S1 demand surge (+60 \%, 7 days, 1/6 of depots) & R1+meter feed (not implemented) & 44.44 & 100.0 & 0.0 & 0.0 \\
S2 unforeseen rail delay (+2 days, 3 days) & MANUAL & 74.16 & 99.81 & 481.6 & 5820.4 \\
S2 unforeseen rail delay (+2 days, 3 days) & R0 & 64.22 & 99.98 & 46.4 & 622.0 \\
S2 unforeseen rail delay (+2 days, 3 days) & R1 & 55.82 & 100.0 & 1.0 & 11.6 \\
S2 unforeseen rail delay (+2 days, 3 days) & R1 two postings & 47.24 & 100.0 & 0.0 & 0.0 \\
S2 unforeseen rail delay (+2 days, 3 days) & R1+meter feed (not implemented) & 44.82 & 100.0 & 0.0 & 0.0 \\
S3 database outage 4 h (day 60, 09:00-13:00) & MANUAL & 73.7 & 99.82 & 423.6 & 5476.8 \\
S3 database outage 4 h (day 60, 09:00-13:00) & R0 & 63.9 & 99.99 & 10.4 & 158.8 \\
S3 database outage 4 h (day 60, 09:00-13:00) & R1 & 55.28 & 100.0 & 0.0 & 0.0 \\
S3 database outage 4 h (day 60, 09:00-13:00) & R1 two postings & 46.56 & 100.0 & 0.0 & 0.0 \\
S3 database outage 4 h (day 60, 09:00-13:00) & R1+meter feed (not implemented) & 44.4 & 100.0 & 0.0 & 0.0 \\
S4 media failure (day 60, 15:00) & MANUAL & 73.7 & 99.82 & 423.6 & 5476.8 \\
S4 media failure (day 60, 15:00) & R0 & 63.68 & 100.0 & 6.8 & 107.4 \\
S4 media failure (day 60, 15:00) & R1 & 55.28 & 100.0 & 0.0 & 0.0 \\
S4 media failure (day 60, 15:00) & R1 two postings & 46.52 & 100.0 & 0.0 & 0.0 \\
S4 media failure (day 60, 15:00) & R1+meter feed (not implemented) & 44.38 & 100.0 & 0.0 & 0.0 \\
\end{longtable}
}
{\footnotesize\noindent The modelled values use a common disruption schedule and are not observations from operating depots. Stock-out depot-product hours, unmet demand and fill rates cover the entire 150-day simulation, including its warm-up/initialization period. No separately computed post-warm-up steady-state outcome is reported; the totals should not be interpreted as such. Source: Authors' own work\par}\medskip

{\footnotesize
\setlength{\tabcolsep}{3pt}
\renewcommand{\arraystretch}{1.15}
\begin{longtable}{@{}L{0.25}L{0.22}L{0.29}L{0.24}@{}}
\caption*{Table S3i. Paired forecast comparison: selected method minus M0}\phantomsection\label{tab:S3i}\\
\toprule
\textbf{Weekly series} & \textbf{MAPE difference (points)} & \textbf{Paired 95\,\% interval (points)} & \textbf{Holm-adjusted p} \\
\midrule
\endfirsthead
\multicolumn{4}{@{}l}{\emph{Table S3i (continued)}}\\
\toprule
\textbf{Series} & \textbf{Difference} & \textbf{Interval} & \textbf{Holm p} \\
\midrule
\endhead
\bottomrule
\endlastfoot
Gasoline demand & $-0.35$ & [$-0.76$, 0.05] & 0.18 \\
Distillate demand & $-0.83$ & [$-1.57$, $-0.11$] & 0.068 \\
Jet demand & 0.03 & [$-0.68$, 0.72] & 0.94 \\
Gasoline stocks & $-4.43$ & [$-6.03$, $-2.87$] & $<0.001$ \\
Distillate stocks & $-5.22$ & [$-6.93$, $-3.58$] & $<0.001$ \\
\end{longtable}
}
{\footnotesize\noindent Differences are computed on the same forecast weeks. Intervals are paired moving-block bootstrap intervals; adjusted p-values use the five-series Holm procedure. Source: Authors' own work\par}\medskip

\subsubsection*{S3.9 Detailed evaluation tables}
{\footnotesize
\setlength{\tabcolsep}{3pt}
\renewcommand{\arraystretch}{1.15}
\begin{longtable}{@{}L{0.17}L{0.15}L{0.17}L{0.17}L{0.17}L{0.17}@{}}
\caption*{Supplementary Table VIII. Integration of multi-source feeds with seeded defects (T1, final validation)}\phantomsection\label{tab:VIII}\\
\toprule
\textbf{Arm} & \textbf{Rows (valid / defective)} & \textbf{Loss of valid rows (retained)} & \textbf{Defective rows accepted} & \textbf{Integration defect rate} & \textbf{Consistent balances} \\
\midrule
\endfirsthead
\multicolumn{6}{@{}l}{\emph{Supplementary Table VIII (continued)}}\\
\toprule
\textbf{Arm} & \textbf{Rows (valid / defective)} & \textbf{Loss of valid rows (retained)} & \textbf{Defective rows accepted} & \textbf{Integration defect rate} & \textbf{Consistent balances} \\
\midrule
\endhead
\bottomrule
\endlastfoot
\multicolumn{6}{@{}l}{\emph{Defect rate 5\,\%, seeds 101 / 102 / 103}}\\
R0 forms & 1,887--2,139 / 95--118 & 0.37 / 0.14 / 0.51\,\% (none) & 62.5 / 64.2 / 56.8\,\% & 3.38 / 2.90 / 3.46\,\% & 100.0 / 98.4 / 97.6\,\% \\
R0 bulk import & same & 40.0 / 35.8 / 48.4\,\% (none) & 44.8 / 55.8 / 43.2\,\% & 40.2 / 36.7 / 48.1\,\% & 98.2 / 0.0 / 96.7\,\% \\
R1 direct & 3,279--3,680 / 172--193 & 0.64 / 0.50 / 0.63\,\% (none) & 7.0 / 8.5 / 5.7\,\% & 0.96 / 0.89 / 0.88\,\% & 89.1 / 86.1 / 87.8\,\% \\
R1 staged & same & 0.58 / 0.36 / 0.52\,\% (all) & 0.0 / 0.0 / 0.0\,\% & 0.55 / 0.34 / 0.49\,\% & 96.4 / 99.2 / 96.7\,\% \\
R1 staged, published scope & as R0 & 0.37 / 0.14 / 0.51\,\% (all) & 0.0 / 0.0 / 0.0\,\% & 0.35 / 0.14 / 0.49\,\% & -- \\
\multicolumn{6}{@{}l}{\emph{Defect rate 10\,\% (seed 104) and 1\,\% (seed 105)}}\\
R0 forms & & 0.46 / 0.05\,\% & 62.3 / 68.2\,\% & 6.55 / 0.74\,\% & 93.3 / 99.2\,\% \\
R1 staged & & 1.07 / 0.05\,\% (all) & 0.3 / 0.0\,\% & 0.99 / 0.05\,\% & 95.8 / 99.2\,\% \\
\end{longtable}
}
{\footnotesize\noindent Notes: R0 rows exclude the outbound deliveries and receipts that the published model cannot represent (1,468--1,616 rows per seed, 42--43\,\% of all rows); the row ``R1 staged, published scope'' restricts R1 to the same four feeds and rows. Balance consistency is measured under each system's own posting rule and is not comparable across systems. Integration defect rate = (valid rows lost + defective rows accepted) / rows. Consistent balances: depot-product stocks equal to the reference computed from the valid rows under each system's posting rule (R0: last count plus orders; R1: ledger); R0 bulk shows 0.0\,\% for seed 102 because the stock-report batches of the last two days were each rejected in full by one defective value, leaving earlier counts in place. Source: Authors' own work\par}\medskip

{\footnotesize
\setlength{\tabcolsep}{3pt}
\renewcommand{\arraystretch}{1.15}
\begin{longtable}{@{}L{0.34}L{0.33}L{0.33}@{}}
\caption*{Supplementary Table IX. Monitoring and alerting on R1 (T2, T5, final validation, SQL Server)}\phantomsection\label{tab:IX}\\
\toprule
\textbf{Measure} & \textbf{Nominal (1$\times$, seed 101)} & \textbf{5$\times$ network, events and clients (seed 102)} \\
\midrule
\endfirsthead
\multicolumn{3}{@{}l}{\emph{Supplementary Table IX (continued)}}\\
\toprule
\textbf{Measure} & \textbf{Nominal (1$\times$, seed 101)} & \textbf{5$\times$ network, events and clients (seed 102)} \\
\midrule
\endhead
\bottomrule
\endlastfoot
Paced window & 3,103 events at 20/s; 1,543 stock changes & 14,016 events at 100/s; 7,668 stock changes \\
Push dashboard latency: median / p95 / max & 0.009--0.010 / 0.016--0.018 / 0.079--0.107 s (4 clients) & 0.007--0.008 / 0.806--1.306 / 7.327--9.061 s (20 clients) \\
Polling dashboard (every 5 s): median / max & 2.96 / 5.00 s & 3.01 / 5.13 s \\
Client observations of stock changes & 6,172 of 6,172 per mode & 153,360 of 153,360 per mode \\
Final dashboard equal to the ledger & 110 / 110 depot-products & 502 / 502 depot-products \\
Expected alerts raised / missed / false positives & 244 / 0 / 0 & 1,144 / 0 / 0 \\
\quad by kind (expected = raised) & reorder 195, critical 10, abnormal 6, count 7, negative book 6, overdue 20 & reorder 909, critical 62, abnormal 30, count 32, negative book 30, overdue 81 \\
Event-driven alert latency: median / max & 0.009 / 0.018 s & 0.007 / 0.415--0.663 s \\
Overdue-delivery alert latency: median / max & 6.4 / 8.1 s (monitor every 10 s) & 8.5 / 8.5 s (81 overdue alerts) \\
\end{longtable}
}
{\footnotesize\noindent Notes: SQL Server final runs. Latency measured from the submission of the triggering transaction (dashboard) or from the alert timestamp written in it (alerts); overdue alerts from the moment the simulated date passed the planned unloading date. Notifications through a Service Broker queue (\citealp{microsoft2025servicebroker}) read by one dispatcher connection, which forwards stock messages to the push clients; each push client then re-reads the changed dashboard rows. Load generator, clients (in four worker processes) and server on one workstation. At 5$\times$, 37,245 of 37,265 push observations above 0.1 s in the tabulated run occurred in the first three tenths of the window; the timing ranges also include the independent repeat. R0 has no refresh, notification or alert function. Source: Authors' own work\par}\medskip

{\footnotesize
\setlength{\tabcolsep}{3pt}
\renewcommand{\arraystretch}{1.15}
\begin{longtable}{@{}L{0.13}L{0.11}L{0.10}L{0.11}L{0.10}L{0.22}L{0.23}@{}}
\caption*{Supplementary Table X. Test MAPE (MASE) one week ahead, EIA weekly data, January 2024 to November 2026}\phantomsection\label{tab:X}\\
\toprule
\textbf{Series} & \textbf{Previous-year mean (M0)} & \textbf{Naive (M1)} & \textbf{Holt-Winters (M4)} & \textbf{Selected method} & \textbf{Selected: MAPE [95\,\% interval] (MASE)} & \textbf{Paired difference from M0, points [95\,\% interval] (Holm $p$)} \\
\midrule
\endfirsthead
\multicolumn{7}{@{}l}{\emph{Supplementary Table X (continued)}}\\
\toprule
\textbf{Series} & \textbf{M0} & \textbf{M1} & \textbf{M4} & \textbf{Selected} & \textbf{Selected: MAPE [interval] (MASE)} & \textbf{Difference from M0 (Holm $p$)} \\
\midrule
\endhead
\bottomrule
\endlastfoot
Gasoline demand & 3.31\,\% (0.75) & 3.74\,\% (0.85) & 3.12\,\% (0.71) & M9 & 2.96\,\% [2.53, 3.42] (0.68) & $-0.35$ [$-0.76$, 0.05] (0.18) \\
Distillate demand & 6.81\,\% (0.82) & 8.82\,\% (1.05) & 7.02\,\% (0.85) & M5 & 5.99\,\% [5.33, 6.75] (0.73) & $-0.83$ [$-1.57$, $-0.11$] (0.068) \\
Jet fuel demand & 6.81\,\% (0.60) & 9.36\,\% (0.82) & 6.79\,\% (0.59) & M5 & 6.84\,\% [6.09, 7.68] (0.60) & 0.03 [$-0.68$, 0.72] (0.94) \\
Gasoline stocks & 5.26\,\% (1.47) & 1.13\,\% (0.31) & 0.93\,\% (0.26) & M5 & 0.83\,\% [0.73, 0.92] (0.23) & $-4.43$ [$-6.03$, $-2.87$] ($<$0.001) \\
Distillate stocks & 6.81\,\% (0.52) & 1.85\,\% (0.14) & 1.98\,\% (0.15) & M5 & 1.59\,\% [1.34, 1.79] (0.12) & $-5.22$ [$-6.93$, $-3.58$] ($<$0.001) \\
\end{longtable}
}
{\footnotesize\noindent Notes: 146 weekly forecast origins for demand and 149 for stocks; methods as in Supplementary Section S3.4 (M5 harmonic regression with ARIMA(1,1,1) errors; M9 Holt-Winters on a training series with the 2020 pandemic weeks replaced by 2019). MASE: mean absolute error divided by the in-sample one-step error of the seasonal naive method over 2010--2020 (\citealp{hyndman2006another}; seasonal form as in \citealp{hyndman2021forecasting}), a scale rather than a test. Selected-method intervals: moving-block bootstrap of that method's MAPE with a fixed block length of 8 weeks (2,000 resamples), unlike the paired differences, whose block length follows the rule of \citet{politis2004automatic}; they describe one method and do not test differences. Paired difference: selected minus M0 on the same weeks, mean difference of the absolute percentage errors with a circular moving-block bootstrap interval; Holm $p$: Diebold--Mariano test with the Harvey--Leybourne--Newbold correction, Holm-adjusted over the five series (Supplementary Table S3i). Source: Authors' own work, from \citet{eia2026weekly}\par}\medskip

{\footnotesize
\setlength{\tabcolsep}{3pt}
\renewcommand{\arraystretch}{1.15}
\begin{longtable}{@{}L{0.15}L{0.17}L{0.14}L{0.17}L{0.2}L{0.17}@{}}
\caption*{Supplementary Table XI. Response time in seconds by load level (T10, SQL Server)}\phantomsection\label{tab:XI}\\
\toprule
\textbf{Load level} & \textbf{R0: general supply forecast} & \textbf{R0: slowest other operation, P95} & \textbf{R1: slowest operation, P95 / maximum} & \textbf{R1: all requests, P95 / P99 / maximum} & \textbf{R1 requests / failures} \\
\midrule
\endfirsthead
\multicolumn{6}{@{}l}{\emph{Supplementary Table XI (continued)}}\\
\toprule
\textbf{Load level} & \textbf{R0: general supply forecast} & \textbf{R0: slowest other operation, P95} & \textbf{R1: slowest operation, P95 / maximum} & \textbf{R1: all requests, P95 / P99 / maximum} & \textbf{R1 requests / failures} \\
\midrule
\endhead
\bottomrule
\endlastfoot
1$\times$ (12 users) & failed 33 of 33 & 0.236 & 0.050 / 0.077 & 0.040 / 0.051 / 0.077 & 697 / 0 \\
2$\times$ (24 users) & failed 37 of 37 & 0.642 & 0.064 / 0.181 & 0.052 / 0.064 / 0.181 & 1,436 / 0 \\
3$\times$ (36 users) & failed 54 of 54 & 1.054 & 0.088 / 0.100 & 0.060 / 0.070 / 0.100 & 2,210 / 0 \\
5$\times$ (60 users) & failed 99 of 99 & 1.309--1.359 & 0.145--0.178 / 0.166--0.210 & 0.095 / 0.128 / 0.285 & 3,509 / 0 \\
5$\times$, R1 without its indexes & -- & -- & 0.160--0.166 / 0.176 & 0.110 / 0.139 / 0.320 & 3,494 / 0 \\
\end{longtable}
}
{\footnotesize\noindent R0: published logic with unique numbering; R1: frozen design; both seed 101, final validation on SQL Server, run on 17 November 2026 with the same calendar (686 days from 1 January 2025). Level $L$ multiplies the network, and with it the volume of its history, and the number of users of the nominal load (emulated users in worker processes of five); statement time-out 60 s; two repetitions of 60 s per run. The 5$\times$ slowest-operation P95 and maximum summarize three runs; the remaining request statistics refer to the tabulated run. P95, P99: 95th and 99th percentiles over all requests of the operation (or of all operations) in both repetitions; the slowest operation is the one with the highest P95. R0 general supply forecast: requests that failed; every failure was the arithmetic overflow of the cross-joined sums (error 8115), none a time-out. Further statistics: Supplementary Table S3d. Source: Authors' own work\par}\medskip

{\footnotesize
\setlength{\tabcolsep}{3pt}
\renewcommand{\arraystretch}{1.15}
\begin{longtable}{@{}L{0.15}L{0.09}L{0.25}L{0.13}L{0.25}L{0.13}@{}}
\caption*{Supplementary Table XII. Seeded-error probes by class (T7, measured, seed 101)}\phantomsection\label{tab:XII}\\
\toprule
\textbf{Class} & \textbf{Probes} & \textbf{R0: prevented / flagged / not enterable / not flagged / not in model} & \textbf{R0 intercepted} & \textbf{R1: prevented / flagged / not enterable / not flagged / not in model} & \textbf{R1 intercepted} \\
\midrule
\endfirsthead
\multicolumn{6}{@{}l}{\emph{Supplementary Table XII (continued)}}\\
\toprule
\textbf{Class} & \textbf{Probes} & \textbf{R0 outcomes} & \textbf{R0 intercepted} & \textbf{R1 outcomes} & \textbf{R1 intercepted} \\
\midrule
\endhead
\bottomrule
\endlastfoot
Invalid & 7 & 3 / 0 / 0 / 4 / 0 & 3/7 & 7 / 0 / 0 / 0 / 0 & 7/7 \\
Cross-record & 16 & 3 / 0 / 0 / 10 / 3 & 3/16 & 8 / 6 / 0 / 2 / 0 & 14/16 \\
Wrong but consistent & 4 & 0 / 0 / 0 / 4 / 0 & 0/4 & 0 / 0 / 0 / 4 / 0 & 0/4 \\
Derived & 3 & 0 / 0 / 1 / 2 / 0 & 1/3 & 0 / 0 / 3 / 0 / 0 & 3/3 \\
All probes & 30 & 6 / 0 / 1 / 20 / 3 & 7/30 & 15 / 6 / 3 / 6 / 0 & 24/30 \\
\end{longtable}
}
{\footnotesize\noindent Notes: Intercepted = prevented, flagged or not enterable. The correct control transaction was accepted and not flagged by both versions. Source: Authors' own work\par}\medskip

{\footnotesize
\setlength{\tabcolsep}{3pt}
\renewcommand{\arraystretch}{1.15}
\begin{longtable}{@{}L{0.22}L{0.12}L{0.1}L{0.11}L{0.1}L{0.08}L{0.15}L{0.12}@{}}
\caption*{Supplementary Table XIII. Mean supply delay of secondary-depot replenishment and its components in hours (T8, modelled, seeds 101--105)}\phantomsection\label{tab:XIII}\\
\toprule
\textbf{Arm} & \textbf{Infor\-mation} & \textbf{Decision} & \textbf{Dispatch wait} & \textbf{Transit} & \textbf{Total} & \textbf{Reduction vs manual, \% (seeds)} & \textbf{Order delay} \\
\midrule
\endfirsthead
\multicolumn{8}{@{}l}{\emph{Supplementary Table XIII (continued)}}\\
\toprule
\textbf{Arm} & \textbf{Infor\-mation} & \textbf{Decision} & \textbf{Dispatch wait} & \textbf{Transit} & \textbf{Total} & \textbf{Reduction vs manual, \% (seeds)} & \textbf{Order delay} \\
\midrule
\endhead
\bottomrule
\endlastfoot
Manual procedure & 24.0 & 6.0 & 10.8 & 33.0 & 73.7 & -- & 29.9 \\
R0 (published logic) & 21.5 & 2.6 & 8.1 & 31.5 & 63.6 & 13.7 (12.7--14.4) & 24.1 \\
R1, deliveries posted at 20:00 & 7.8 & 13.3 & 2.0 & 32.1 & 55.3 & 25.0 (24.4--25.8) & 21.2 \\
R1, deliveries posted at 11:00 and 20:00 & 4.1 & 9.0 & 2.3 & 31.2 & 46.5 & 36.9 (36.0--38.3) & 13.1 \\
R1 with an hourly meter feed (not implemented) & 0.0 & 5.0 & 6.7 & 32.7 & 44.3 & 39.8 (38.8--41.5) & 5.0 \\
\end{longtable}
}
{\footnotesize\noindent Supply delay: from the moment the inventory position falls below the reorder point to the arrival of the replenishment; order delay: to the issued order. Means summarize the specified model horizon; stock-out totals reported separately include the warm-up period. Components are means over all replenishments and may not add exactly to the total because of rounding. Transit times follow the same distributions in every arm; their means differ slightly with the mix of rail and road orders. The hypothetical hourly-meter scenario yielded a mean dispatch wait of 6.7 h, compared with 2.0--2.3 h in the other R1 scenarios; these values are outputs of the discrete-event model, which does not isolate the contribution of order-submission timing and loading-slot cut-offs. Source: Authors' own work\par}\medskip

{\footnotesize
\setlength{\tabcolsep}{3pt}
\renewcommand{\arraystretch}{1.15}
\begin{longtable}{@{}L{0.26}L{0.29}L{0.29}L{0.16}@{}}
\caption*{Supplementary Table S3j. Summary of results: published logic or manual procedure and revised artefact}\phantomsection\label{tab:S3j}\\
\toprule
\textbf{Measure} & \textbf{Published logic (R0) or manual procedure} & \textbf{Revised artefact (R1)} & \textbf{Evidence} \\
\midrule
\endfirsthead
\multicolumn{4}{@{}l}{\emph{Supplementary Table S3j (continued)}}\\
\toprule
\textbf{Measure} & \textbf{Published logic (R0) or manual procedure} & \textbf{Revised artefact (R1)} & \textbf{Evidence} \\
\midrule
\endhead
\bottomrule
\endlastfoot

Integration defect rate of multi-source feeds (lost valid + accepted defective rows / rows; 5\,\% seeded defects)
& R0 form entry 2.90--3.46\,\% of the four feeds it can represent (42--43\,\% of all rows not representable); R0 bulk import 36.7--48.1\,\%
& R1 staged, same four feeds 0.14--0.49\,\%; all six feeds 0.34--0.55\,\%
& Measured \\

Production records and orders accepted over one year of nominal operation
& R0 3,009/12,351 (24.4\,\%)
& R1 12,351/12,351 (100.0\,\%)
& Measured \\

Report fields equal to independent reference values (eight views with the same fields in both versions; the general supply forecast separately)
& R0 logic 4,697/4,733 (99.24\,\%) (nominal network), 6,875/6,941 (99.05\,\%) (2$\times$ network); general supply forecast fails (arithmetic overflow, error 8115); published end to end 1,105/4,861 (22.73\,\%) of all fields
& R1 4,733/4,733 (100.00\,\%) and 6,941/6,941 (100.00\,\%); general supply forecast 50/50
& Measured \\

Maximum report generation time
& R0 logic: general supply forecast fails after up to 1.4 s (nominal) and 3.9 s (2$\times$), isolated T3 reports (T10 concurrent nominal: 3.6 s); other reports at most 0.019 s and 0.025 s
& R1 0.037 s (nominal), 0.049 s (2$\times$)
& Measured \\

Display latency of committed stock changes (push dashboard: median / max)
& R0: no refresh or notification (not measurable)
& R1 median 0.009--0.010 s, maximum 0.079--0.107 s (1$\times$); median 0.007--0.008 s and maximum 7.327--9.061 s (5$\times$); 5-s polling max 5.13 s
& Measured \\

Expected alerts raised (false positives); notification latency
& R0: no alert function (not measurable)
& R1 244/244 (1$\times$) and 1,144/1,144 (5$\times$), 0 false positives; event-driven max 0.415--0.663 s, overdue deliveries max 8.5 s (monitor every 10 s)
& Measured \\

Test MAPE one week ahead, EIA weekly series 2024--2026 (gasoline / distillate / jet demand; gasoline / distillate stocks)
& Previous-year mean of the published views (M0) 3.31 / 6.81 / 6.81 / 5.26 / 6.81\,\%
& Method selected on validation 2.96 / 5.99 / 6.84 / 0.83 / 1.59\,\%; paired difference from M0 established for gasoline stocks and distillate stocks (Holm-adjusted $p<0.001$, $p<0.001$), not for gasoline demand, distillate demand, jet demand (Holm-adjusted $p=0.18$, $p=0.068$, $p=0.94$)
& Measured \\

Seeded errors intercepted (prevented, flagged or not enterable); probes both versions can represent, and all probes
& R0 7/27
& R1 21/27; all probes 24/30
& Measured \\

Business actions per operating day
& Manual procedure 1,670.6; R0 1,290.7 ($-$22.7\,\%)
& R1 526.0 ($-$68.5\,\%)
& Modelled \\

Remaining errors (reduction relative to the manual procedure; mean and 2.5--97.5 percentiles over 20 Monte Carlo replications)
& R0 $-$57.4\,\% [52.6, 64.0]
& R1 $-$93.1\,\% [91.8, 94.4]
& Modelled \\

Mean supply delay, reorder point crossed to arrival (seeds 101--105)
& Manual procedure 73.7 h; R0 63.6 h ($-$13.7\,\%)
& R1 55.3 h ($-$25.0\,\%); deliveries also posted at 11:00 46.5 h ($-$36.9\,\%, seeds 36.0--38.3\,\%)
& Modelled \\

Forbidden role--operation attempts that succeeded
& R0 25/25 through the application login (7/25 reachable from the menus)
& R1 0/36, of which 0/25 on the pairs the published model can express (permitted operations 32/32)
& Measured \\

Response time at five times the nominal load (seconds; 95th percentile of the slowest operation, 99th percentile and maximum of all requests)
& R0 logic: general supply forecast failed in 99 of 99 requests (arithmetic overflow, error 8115; 0 time-outs); other operations highest P95 1.309--1.359 s
& R1: slowest operation (general supply forecast) P95 0.145--0.178 s; all requests P99 0.128 s, maximum 0.285 s (0 failures in 3,509 requests); without its indexes, slowest operation P95 0.160--0.166 s
& Measured \\

Data lost after a simulated media failure (three trials)
& R0 strategy (full backup at the start of the day): all documents of the day (40, 40, 40 in the three trials; 67--150 committed transactions)
& R1 (full backup + log backups every 60 s): 58--106 rows committed after the last log backup, at most 32 stock movements; documents lost 30, 0, 0 in the three trials (the first trial lost 30 of that day's 40 documents, all within the 60-s log-backup interval in the paced replay)
& Measured \\

Unaided task completion by operators
& Not tested (no participants)
& Not tested (no participants)
& Not evaluated \\

\end{longtable}
}

{\footnotesize\noindent
Note: Results are classified as measured or modelled.
Comparisons are restricted to capabilities shared by both implementations;
additional R1 capabilities are reported separately.
Source: Authors' own work\par}\medskip

\subsection*{S4. Evaluation settings and sensitivity checks}\phantomsection\label{sec:S4}
The numerical results in the main article refer to the evaluated final configuration and to the scenario variants stated in each table. The following checks clarify the scope of the comparisons; they do not refer to separate deposited experiment files.

{\footnotesize
\setlength{\tabcolsep}{3pt}
\renewcommand{\arraystretch}{1.15}
\begin{longtable}{@{}L{0.18}L{0.42}L{0.40}@{}}
\caption*{Table S4. Evaluation settings and sensitivity checks}\phantomsection\label{tab:S4}\\
\toprule
\textbf{Scenario} & \textbf{Evaluated setting} & \textbf{Interpretation} \\
\midrule
\endfirsthead
\multicolumn{3}{@{}l}{\emph{Table S4 (continued)}}\\
\toprule
\textbf{Scenario} & \textbf{Evaluated setting} & \textbf{Interpretation} \\
\midrule
\endhead
\bottomrule
\endlastfoot
T4 & Rolling-origin validation followed by evaluation on a held-out public-data test interval & Model selection and held-out assessment are distinct \\
T8 & Outbound deliveries posted at 20:00 versus twice daily at 11:00 and 20:00 & Discrete-event simulation variants, not field measurements; mean delays 55.3 versus 46.5 h \\
T8 sensitivity & One-at-a-time parameter and reporting-schedule changes on seeds 101--103 only & A three-seed sensitivity subset of the five-seed main evaluation; ranges are conditional and are not five-seed uncertainty intervals \\
T12 stock-outs & Entire 150-day simulation, including the warm-up/initialization period & Cumulative stock-out hours and fill rates are not post-warm-up steady-state estimates \\
T10 & R1 evaluated with and without its indexes under fivefold load & Slowest-operation P95 0.145--0.178 versus 0.160--0.166 s in the reported measurements \\
\end{longtable}
}
\tsrc

\subsection*{S5. Experimental specifications and scope}\phantomsection\label{sec:S5}
\begin{itemize}[label=--,leftmargin=*,itemsep=2pt,topsep=4pt]
\item \textbf{Software environment.} Microsoft SQL Server 2022 Developer, version 16.0.4325.4, on Linux with a 2\,GB memory limit; twelve logical processors on the workstation running the Python client harness. SQL Server stores the reconstructed R0 tables and the R1 implementation. The missing Delphi form resources prevent execution of the original graphical client.
\item \textbf{Generator and validation.} Development seed 1; independent synthetic validation seeds 101--105 for the indicated main analyses, but seeds 101--103 for the T8 one-at-a-time sensitivity checks; one-, two-, three- and fivefold load as described in T10. The test run calendar and data scope are specified in the main results. Input records, source references and probe labels are generated before the associated test.
\item \textbf{Forecasting series.} Public U.S. weekly product-supplied and stock observations from the U.S. Energy Information Administration; validation through 2023 and held-out test observations from January 2024 through November 2026. All comparisons preserve the corresponding forecast-origin dates.
\item \textbf{Modelled outcomes.} Process-action rates and error probabilities, replenishment timing and physical transport distributions are model parameters; the resulting quantities are not field performance measurements.
\end{itemize}

\clearpage
\bibliographystyle{plainnat}
\bibliography{references}
\end{document}